\documentclass[letterpaper,twocolumn,10pt]{article}
\usepackage{usenix,epsfig}
\usepackage{epsfig}
\usepackage{booktabs} % For formal tables
\usepackage{tabularx}
\usepackage{multirow}
\usepackage{mathtools}
\usepackage{nicefrac}
\usepackage{amsfonts,amssymb,amsthm}
\usepackage{url}
\usepackage{commath}
\usepackage{algorithm}
\usepackage[inline]{enumitem}
\usepackage[noend]{algpseudocode}

\DeclarePairedDelimiter\floor{\lfloor}{\rfloor}
\usepackage{titling}
\usepackage{caption}
\usepackage{subcaption}
\begin{document}
\date{}
\title{\Large \bf Accelerating PageRank using Partition-Centric Processing}
	\vspace{-1em}
\author{%
% author names are typeset in 11pt, which is the default size in the author block
{Kartik Lakhotia, Rajgopal Kannan, Viktor Prasanna}\vspace{1mm}\\%
% add some space between author names and affils
%\fontsize{10}{10}\selectfont\itshape
% 20080211 CAUSAL PRODUCTIONS
% separate superscript on following line from affiliation using narrow space
Ming Hsieh Department of Electrical Engineering, University of Southern California\\
%\fontsize{9}{9}\selectfont\ttfamily\upshape
%
% 20080211 CAUSAL PRODUCTIONS
% in the following email addresses, separate the superscript from the email address 
% using a narrow space \,
% the reason is that Acrobat Reader has an option to auto-detect urls and email
% addresses, and make them 'hot'.  Without a narrow space, the superscript is included
% in the email address and corrupts it.
% Also, removed ~ from pre-superscript since it does not seem to serve any purpose
\{klakhoti, rajgopak, prasanna\}@usc.edu
 %\vspace{-3.5mm}
% add some space between email and affil
}
\maketitle

%\begin{abstract}
\subsection*{Abstract}
	PageRank is a fundamental link analysis algorithm that also functions as a key representative of the performance of Sparse Matrix-Vector~(SpMV) multiplication. The traditional PageRank implementation generates fine granularity random memory accesses resulting in large amount of wasteful DRAM traffic and poor bandwidth utilization. In this paper, we present a novel Partition-Centric Processing Methodology~(PCPM) to compute PageRank, that drastically reduces the amount of DRAM communication while achieving high sustained memory bandwidth. PCPM uses a Partition-centric abstraction coupled with the Gather-Apply-Scatter~(GAS) programming model. By carefully examining how a PCPM based implementation impacts communication characteristics of the algorithm, we propose several system optimizations that improve the execution time substantially. More specifically, we develop~\begin{enumerate*}[label={(\arabic*)}]
		\item a new data layout that significantly reduces communication and random DRAM accesses, and
		\item branch avoidance mechanisms to get rid of unpredictable data-dependent branches.
	\end{enumerate*}  
	%The state-of-the-art Vertex-centric PageRank implementation also uses GAS model but suffers from random DRAM accesses and redundant update value propagation from nodes to their neighbors. In contrast, our Partition-centric abstraction enables streaming memory accesses with at most single update transfer from a node to a partition, thereby decreasing random accesses and redundancy effectively. We also develop a novel bipartite Partition-Node Graph~(PNG) data layout that enables streaming memory accesses in PCPM with small pre-processing overhead.
	
	We perform detailed analytical and experimental evaluation of our approach using 6 large graphs and demonstrate an average $2.7\times$ speedup in execution time and $1.7\times$ reduction in communication volume, compared to the state-of-the-art. We also show that unlike other GAS based implementations, PCPM is able to further reduce main memory traffic by taking advantage of intelligent node labeling that enhances locality. Although we use PageRank as the target application in this paper, our approach can be applied to generic SpMV computation.
%\end{abstract}
%\vspace{-1mm}
\section{Introduction}\label{sec:intro}
%\begin{itemize}
%	\item Graph Analytics, large scale and challenges- memory wall etc.
%	\item Pagerank algorithm description, use case, random access to vertices, possibly a graph that gives a comparison of lower bound on memory communication vs total communication or the amount of cycles stalled on memory or cache misses etc.
%	\item Briefly mention that recent works have tried to split computation into two phases to improve spatial locality but suffer from 2 issues
%	\item In this paper, we present a novel technique to reduce the redundancies and a novel data layout/structure to achieve streaming memory performance
%	\item Major contributions are: list them
%	\item Rest of the paper is divided as follows
%\end{itemize}
Graphs are the preferred choice of data representation in many fields such as web and social network analysis~\cite{graphinweb, barabasi, wattsstrogatz, tao}, biology~\cite{biology}, transportation~\cite{osm, applications} etc. The growing scale of problems in these areas has generated substantial research interest in high performance graph analytics. A large fraction of this research is focused on shared memory platforms because of their low communication overhead compared to distributed systems~\cite{cost}. High DRAM capacity in modern systems further allows in-memory processing of large graphs on a single server~\cite{ligra, grace,graphmat}. However, efficient utilization of compute power is challenging even on a single node because of the \begin{enumerate*}[label={(\arabic*)}]
	\item low computation-to-communication ratio and,
	\item irregular memory access patterns of graph algorithms.
\end{enumerate*} The growing disparity between CPU speed and DRAM bandwidth, termed memory wall~\cite{wall}, has become a key issue in high performance graph analytics.

PageRank is a quintessential algorithm that exemplifies the performance challenges posed by graph computations. It 
%is also a fundamental link analysis algorithm~\cite{pagerank} that 
iteratively performs Sparse Matrix-Vector~(SpMV) multiplication over the adjacency matrix of the target graph and the current PageRank vector $\overrightarrow{PR}$ to generate new PageRank values. The irregularity in adjacency matrices leads to random accesses to $\overrightarrow{PR}$ with poor \textit{spatial} and \textit{temporal} locality. The resulting cache misses and communication volume become the performance bottleneck for PageRank computation. Since many graph algorithms can be similarly modeled as a series of SpMV operations~\cite{graphmat}, optimizations on PageRank can be easily generalized to other algorithms. 
% Many graph algorithms can be modeled as a series of SpMV computations over adjacency matrix, with similar data access patterns and challenges as PageRank~\cite{graphmat}. Hence, performance optimizations over PageRank can be easily generalized to other graph algorithms as well.

%In this paper, we present a novel Partition-Centric programming model~(PCPM) that substantially reduces the amount of DRAM communication and avoids random accesses to main memory to achieve high bandwidth. Although we demonstrate the advantages of our approach on PageRank in this paper, it can be applied for any SpMV based algorithms.
%\begin{figure}
%	\centering
%	\includegraphics[width=0.9\linewidth]{figures/rankValueTraffic.pdf}
%	\caption{DRAM traffic generated from PageRank vector accesses. It constitutes a large fraction of total communication in the algorithm.}
%	\label{fig:vvTraffic}
%\end{figure}

Recent works have proposed the use of Gather-Apply-Scatter~(GAS) model to improve locality and reduce communication for SpMV and PageRank~\cite{shijie, spmv, propagationBlocking}. This model splits computation into two phases: \textit{scatter} current source node values on edges and \textit{gather} propagated values on edges to compute new values for destination nodes. The $2$-phased approach restricts access to either the current $\overrightarrow{PR}$ or new $\overrightarrow{PR}$ at a time. This provides opportunities for cache-efficient and lock-free parallelization of the algorithm.
%Specifically, storing the scattered values into bins corresponding to the destination nodes enhances locality and cache performance, resulting in overall reduction in the amount of data communicated with DRAM.
%~\cite{spmv, propagationBlocking} efficiently parallelize computation within each phase without the need of locks or atomics.

We observe that although this approach exhibits several attractive features, it also has some drawbacks leading to inefficient memory accesses, both quantitative as well as qualitative. First, we note that while scattering, a vertex {\it repeatedly} writes its value on all outgoing edges, resulting in large number of reads and writes. We also observe that the Vertex-centric graph traversal in~\cite{spmv,propagationBlocking} results in {\it random} DRAM accesses and the Edge-centric traversal in~\cite{xstream,shijie} scans edge list in coordinate format which increases the number of reads. 

Our premise is that by changing the focus of computation from a single vertex or edge to a {\it cacheable} group of vertices (partition), we can effectively identify and reduce redundant edge traversals as well as avoid random accesses to DRAM, while still retaining the benefits of GAS model.  Based on these insights, we develop a new {\it Partition-Centric} approach to compute PageRank. The major contributions of our work are:
%\vspace{-0.2mm}
\begin{enumerate}
	\itemsep-0.2mm
	\item We propose a Partition-Centric Processing Methodology~(PCPM) that propagates updates from nodes to partitions and reduces the redundancy associated with GAS model.
	\item By carefully evaluating how a PCPM based implementation impacts algorithm behavior, we develop several system optimizations that substantially accelerate the computation, namely, \begin{enumerate*}[label={(\alph*)}]
		\item a new data layout that drastically reduces communication and random memory accesses,
		\item branch avoidance mechanisms to remove unpredictable branches.
		\end{enumerate*}
	%We show that generating PNG involves very low cost pre-processing that does not impose any significant overhead on the PageRank computation.
	\item We demonstrate that PCPM can take advantage of intelligent node labeling to further reduce the communication volume. Thus, PCPM is suitable even for high locality graphs.
	\item We conduct extensive analytical and experimental evaluation of our approach using 6 large datasets. On a 16-core shared memory system, PCPM achieves $2.1$$\times$$-3.8\times$ speedup in execution time and $1.3$$\times$$-2.5\times$ reduction in main memory communication over state-of-the-art.
	\item We show that PCPM can be easily extended to weighted graphs and generic SpMV computation~(section~\ref{sec:ext}) even though it is described in the context of PageRank algorithm in this paper. 
\end{enumerate}
%The rest of the paper is organized as follows: Section~\ref{sec:background} covers background and related work;  Section~\ref{sec:pcpm} describes PCPM along with PNG data layout; Section~\ref{sec:analyticEval} gives detailed analytical evaluation and comparison of PCPM with the state-of-the-art; Section~\ref{sec:expEval} reports the experimental results and Section~\ref{sec:conclusion} concludes the paper.
%GAS, which is a widely used programming model in graph analytics domain, works by %splitting the computation in 2 phases: \begin{enumerate}
%	\item scatter updates on outgoing edges
%	\item gather incoming updates
%\end{enumerate}~\cite{\begin{comment}pregel, power graph\end{comment}}. By partitioning the graph and writing updates sequentially into the destination partition, the GAS model ensures full cache line utilization and thereby, reduced memory traffic~\cite{\begin{comment}XStream\end{comment}}.
%\vspace{-2mm}
\section{Background and Related Work}\label{sec:background}%\vspace{-0.5mm}
\subsection{PageRank Computation}\label{sub:PR}
In this section, we describe how PageRank is calculated and what makes it challenging for the conventional implementation to achieve high performance. Table~\ref{table:notation} lists a set of notations that we use to mathematically represent the algorithm. 
\begin{table}[htbp]
	%\vspace{-1mm}
	\centering
	\caption{List of graph notations}
	\label{table:notation}
	\resizebox{\linewidth}{!}{%	
	\begin{tabular}{|c|c|}
		\hline
		$G(V,E)$                 & Input directed graph                                              \\ \hline
%		$V$                      & vertex set of $G(V,E)$                                            \\ \hline
%		$E$                      & edge set of $G(V,E)$                                              \\ \hline
		$A$						 & adjacency matrix of $G(V,E)$ \\ \hline
		$N_i(v)$                 & in-neighbors of vertex $v$                                        \\ \hline
		$N_o(v)$                 & out-neighbors of vertex $v$                                       \\ \hline
		$\overrightarrow{PR}_i$  & PageRank value vector after $i^{th}$ iteration                    \\ \hline
		$\overrightarrow{SPR}$ & scaled PageRank vector $\big(SPR(v) = \frac{PR_i(v)}{\abs{N_o(v)}}\big)$ \\ \hline
		$d$						 & damping factor in PageRank algorithm \\ \hline
		\end{tabular}
	}
	%\vspace{-2mm}
\end{table}

PageRank is computed iteratively. In each iteration, all vertex values are updated by the new weighted sum of their in-neighbors' PageRank, as shown in equation~\ref{eq:PR}. 
\begin{equation} \label{eq:PR}
\begin{aligned}
PR_{i+1}(v) = \frac{1-d}{\abs{V}}\ +\ d\sum_{u\in N_{i}(v)}\frac{PR_{i}(u)}{\abs{N_{o}(u)}}
\end{aligned}
\end{equation}
%From a linear algebraic perspective equation~\ref{eq:PR} translates to computing multiplication of sparse adjacency matrix $A^T$ with the scaled PageRank vector $\overrightarrow{SPR}$.

To assist visualization of some techniques and optimizations, we also use the Linear Algebraic perspective where a PageRank iteration can be re-written as follows:
\begin{equation}\label{eq:PRLA}
\overrightarrow{PR}_{i+1} = \frac{1-d}{\abs{V}}\cdot\boldsymbol{1}_{\abs{V}\times1} + A^{T}\cdot\overrightarrow{SPR}_{i}
\end{equation}
The most computationally intensive term that dictates the performance of computing this equation is the Sparse Matrix-Vector~(SpMV) multiplication $A^T\cdot\overrightarrow{SPR}_i$. Henceforth, we will focus on the SpMV term to improve the performance of PageRank algorithm. 

PageRank is typically computed in pull direction~\cite{ligra, gunrock, graphmat, galois} where each vertex pulls the value of its in-neighbors and accumulates into its own value, as shown in algorithm~\ref{alg:PRPull}. This corresponds to traversing $A$ in a column-major order and computing the dot product of each column with the scaled PageRank vector $\overrightarrow{SPR}$. 
%Since the iterative SpMV computation uses only $\overrightarrow{SPR}$, we use the $PR[]$ array to store $\overrightarrow{SPR}$ during the computation and $\overrightarrow{PR}$ while outputting final results. 
%as well~(alg.~\ref{alg:PRPull}). 
%\vspace{-1mm}
\begin{algorithm}
	\caption{Pull Direction PageRank~(PDPR) Iteration}
	\label{alg:PRPull}
	\begin{algorithmic}[1]
%		\Function{PageRank}{$G(V,E), PR[]$}
		\For {$v\in V$}
		\State {$temp = 0$}
		\ForAll {$u\in N_i(v)$}
		\State {$temp += PR[u]$}
		\EndFor
		\State {$PR_{next}[v] = \frac{(1-d)\times\abs{V}^{-1}\ +\ d\times temp}{\abs{N_o(v)}}$	}	
		\EndFor
		\State {swap$(PR, PR_{next})$}
%		\EndFunction
	\end{algorithmic}
\end{algorithm}
 
In the pull direction implementation, each column completely owns the computation of the corresponding element in the output vector. This enables all columns of $A$ to be traversed asynchronously in parallel without the need to store partial sums in memory. On the contrary, in the push direction, each node updates its out-neighbors by adding its own value to them. This requires a row-major traversal of $A$ and storage for partial sums since each row contributes partially to multiple elements in the output vector. Further, synchronization is needed to ensure conflict-free processing of multiple rows that update the same output element.

\begin{figure}
	\centering
	\includegraphics[width=0.9\linewidth]{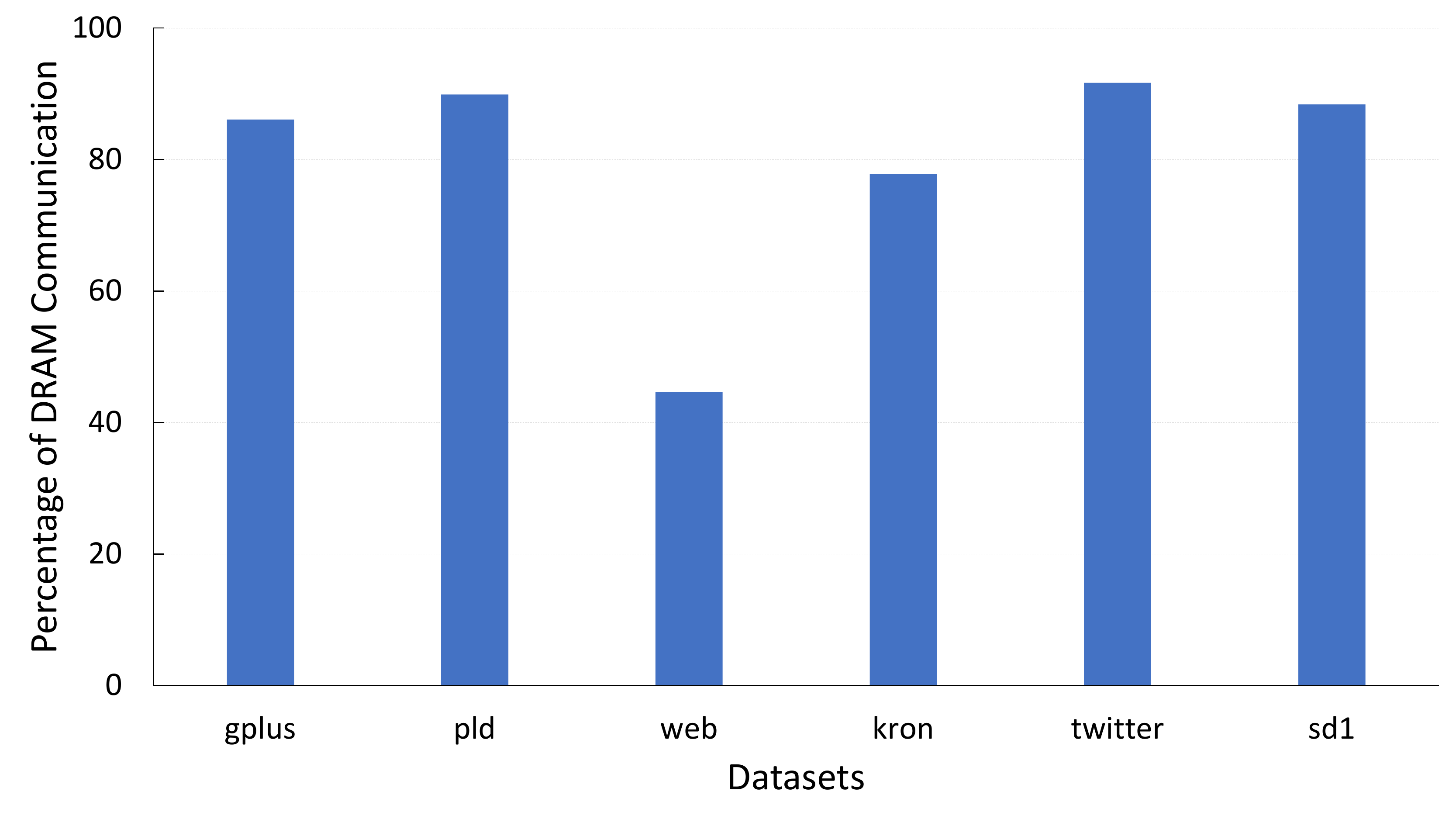}
	\caption{Percentage contribution of vertex value accesses to the total DRAM traffic in a PageRank iteration.}
	\label{fig:percentageTraffic}
	%\vspace{-2mm}
\end{figure}

\begin{figure}
	\centering
	\includegraphics[width=\linewidth]{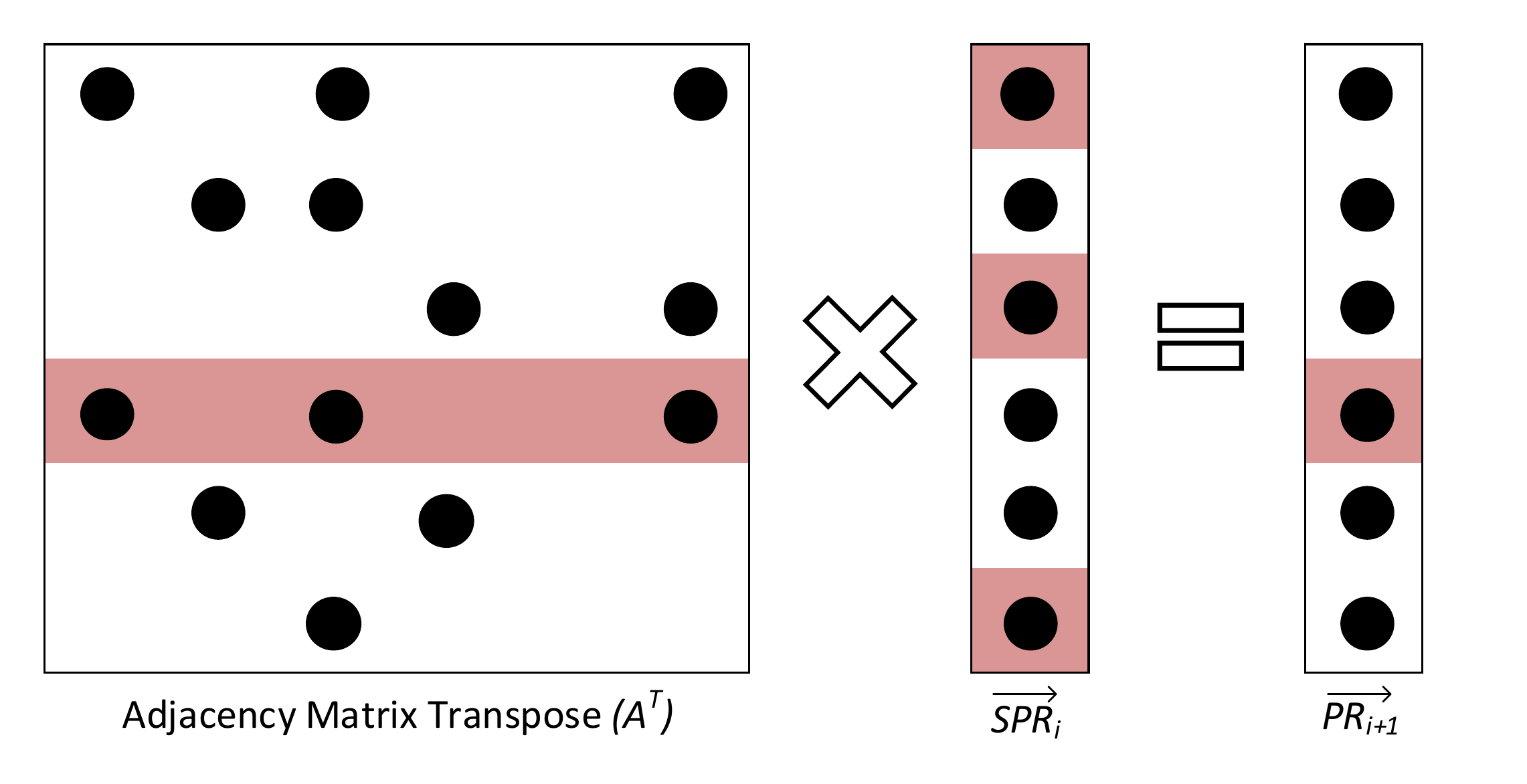}
	\caption{Access locations into $\protect\overrightarrow{SPR}$ are derived from non-zero column indices in rows of $A^T$ and tend to be highly irregular.}
	\label{fig:spmv}
\end{figure}

\noindent\textbf{Performance Challenges:} Sparse matrix layouts like Compressed Sparse Column~(CSC) store all non-zero elements of a column sequentially in memory allowing fast column-major traversal of~$A$~\cite{boost}. However, as shown in fig.~\ref{fig:spmv}, the neighbors of a node~(non-zero columns in adjacency matrix) can be scattered anywhere in the graph and reading their values results in random accesses~(single or double word) to $\overrightarrow{SPR}$ in pull direction computation. Similarly, the push direction implementation uses a Compressed Sparse Row~(CSR) format for fast row-major traversal of $A$ but suffers from random accesses to the partial sums vector. These low locality and fine granularity accesses incur high cache miss ratio and contribute a large fraction to the overall memory traffic as shown in fig.~\ref{fig:percentageTraffic}.

%\vspace{-1mm}
\subsection{Related Work}
The performance of PageRank depends heavily on the locality in memory access patterns of the graph~(which we refer to as \textit{graph locality}). Since node labeling has significant impact on graph locality, many prior works have investigated the use of node reordering or clustering~\cite{permuting, rcm, webbase, metis} to improve the performance of graph algorithms.
Reordering based on spatial and temporal locality aware placement of neighbors~\cite{gorder, recall} has been shown to further outperform the well known clustering and tree-based techniques.
% In~\cite{hipc}, we extended this line of research to propose a Block Reordering algorithm that performs joint spatio-temporal locality optimization~\cite{hipc}. 
Such sophisticated algorithms provide significant speedup but also introduce substantial pre-processing overhead which limits their practicability. 
%Other approaches like~\cite{fillingCurves} and~\cite{cost} propose the use of Hilbert curves to increase locality by reordering edges in the graph. However, this imposes a specific order for edge traversal and makes it challenging to parallelize the application. 
In addition, scale-free graphs like social networks are less tractable by reordering transformations because of their skewed degree distribution.

Cache Blocking~(CB) is another technique used to accelerate graph processing~\cite{cacheBlocking,floydWarshall,gridgraph}. CB attempts to induce locality by restricting range of randomly accessed nodes and has been shown to reduce cache misses~\cite{everything}. CB partitions $A$ along rows, columns or both into multiple block matrices. Each block matrix can be stored in CSR or coordinate~(COO) format. However, SpMV computation with CB requires the partial sums to be re-read for each block. The extremely sparse nature of these block matrices also reduces the locality in partial sum accesses~\cite{cbUtility}.

Gather-Apply-Scatter~(GAS) is another popular model incorporated in many graph analytics frameworks ~\cite{pregel,xstream,powergraph}. It splits the analytic computation into \textit{scatter} and \textit{gather} phases. In the \textit{scatter} phase, source vertices transmit updates on all of their outgoing edges and in the \textit{gather} phase, these updates are processed to compute new values for corresponding destination vertices. The \textit{updates} for PageRank algorithm correspond to \textit{scaled PageRank values} defined earlier in section~\ref{sub:PR}. 

Binning exploits the $2$-phased computation model by storing the updates in a semi-sorted manner. This induces \textit{spatio-temporal} locality in access patterns of the algorithm. Binning can be used in conjunction with both Vertex-centric or Edge-centric paradigms. Zhou et al.~\cite{shijie,hpec} use a custom sorted edge list with Edge-centric processing to reduce DRAM row activations and improve memory performance. However, their sorting mechanism introduces a non-trivial pre-processing cost and imposes the use of COO format. This results in larger communication volume and execution time than the CSR based Vertex-centric implementations~\cite{propagationBlocking,spmv}. 

GAS model is also \textit{inherently sub-optimal} when used with either Vertex-centric or Edge-centric abstractions. This is because it traverses the entire graph twice in each iteration. Nevertheless, Binning with Vertex-centric GAS~(BVGAS) is the state-of-the-art methodology on shared memory platforms~\cite{propagationBlocking, spmv} and we use it as baseline for comparison in this paper.
\section{Partition-Centric Processing}\label{sec:pcpm}
We propose a new Partition-Centric Processing Methodology~(PCPM) that significantly improves the efficiency of processor-memory communication over that achievable with current Vertex-centric or Edge-centric methods. We define \textit{partitions} as disjoint sets of contiguously labeled nodes. The Partition-Centric abstraction then perceives the graph as a set of links from each node to the partitions corresponding to the neighbors of the node. We use this abstraction in conjunction with the 2-phased Gather-Apply-Scatter~(GAS) model.

During the PCPM scatter phase, each thread processes one partition at a time. Processing a partition $p$ means propagating messages from nodes in $p$ to the neighboring partitions. A message to a partition $p'$ comprises of the update value of source node~($PR[v]$) and the list of out-neighbors of $v$ that lie in $p'$. PCPM caches the vertex data of $p$ and streams the messages to the main memory. The messages from $p$ are generated in a Partition-centric manner i.e. messages from all nodes in $p$ to a neighboring partition $p'$ are generated consecutively and are not interleaved with messages to any other partition. 

During the gather phase, each thread scans all messages destined to one partition $p$ at a time. A message scan applies the update value to all nodes in the neighbor list of that message. Partial sums of nodes in $p$ are cached and messages are streamed from the main memory. After all messages to $p$ are scanned, the partial sums~(new PageRank values) are written back to DRAM.
%In other words, any vertex sends at most $1$ update to a given partition along with a list of neighbors that lie in that partition. 
%As we will see later, the neighbor list is only communicated once~(in the first iteration or during pre-processing). This prevents PCPM from traversing the entire graph in scatter phase and drastically reduces the amount of DRAM data transfer.

With static pre-allocation of distinct memory spaces for each partition to write messages, PCPM can asynchronously scatter or gather multiple partitions in parallel. In this section, we provide a detailed discussion on PCPM based computation and the required data layout.
%\vspace{-2mm}
\subsection{Graph Partitioning}
\begin{figure}[htbp]
%	\vspace{-2mm}
	\centering
	\begin{subfigure}{0.9\linewidth}
		\centering
		\includegraphics[width=\linewidth]{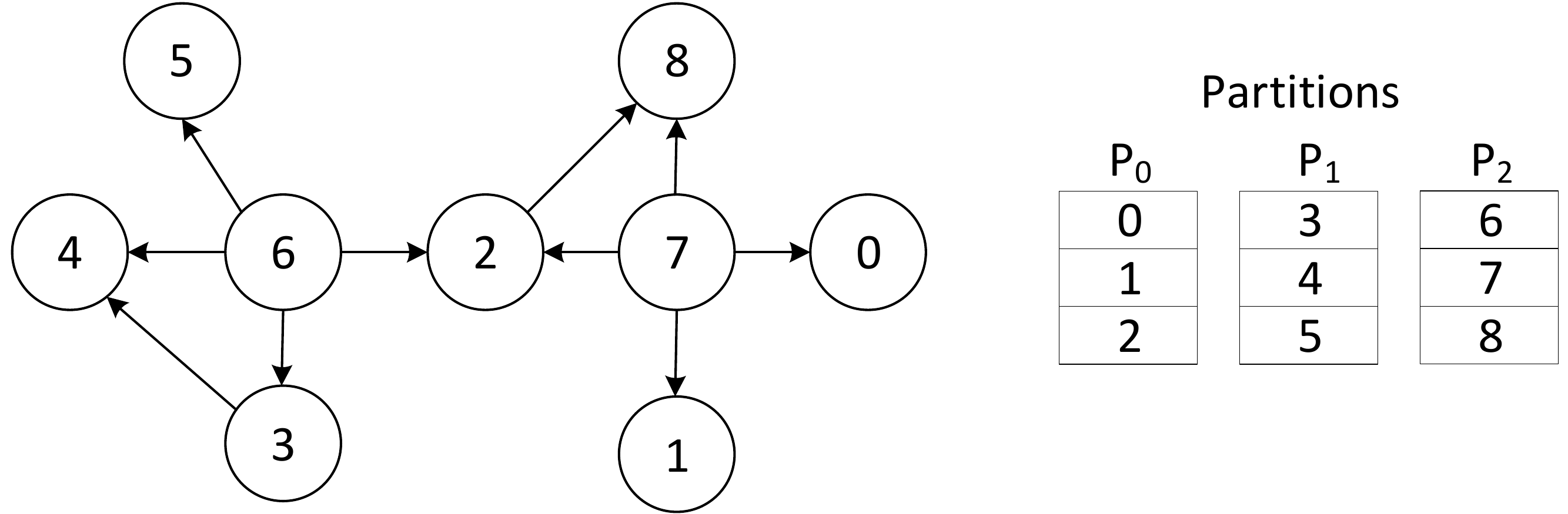}
		\caption{Example graph with partitions of size $3$}
		\label{fig:graphEx}
	\end{subfigure}
	%\vspace{1mm}
	\begin{subfigure}{\linewidth}
%		\vspace{-1mm}
		\centering
		\includegraphics[width=0.9\linewidth]{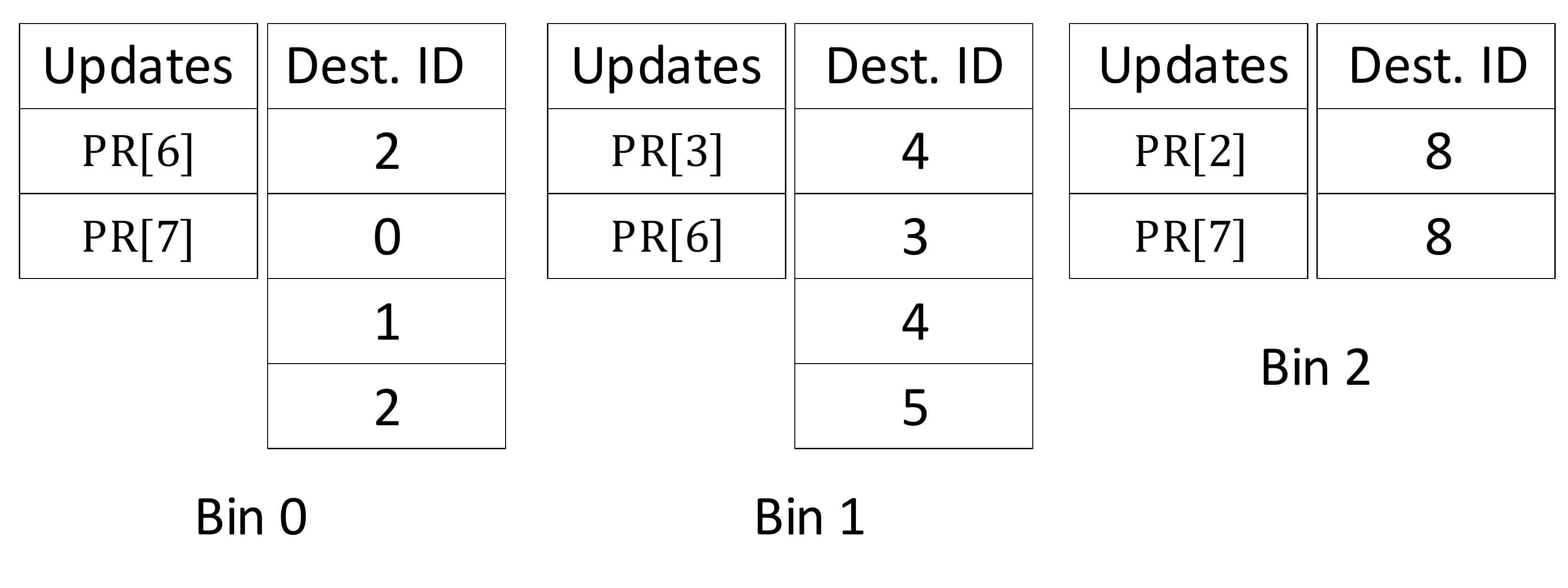}
%		\vspace{-40pt}
		\caption{Bins store update value and list of destination nodes}
		\label{fig:graphBins}
	\end{subfigure}
	\caption{Graph Partitioning and messages inserted in bins during scatter phase}
	\label{fig:partitioning}
%	\vspace{-3mm}
\end{figure}

We employ a simple approach to divide the vertex set $V$ into partitions. We create equisized partitions of size $q$ where partition $P_i$ owns all the vertices with index $\in[i*q, (i+1)*q)$ as shown in fig.~\ref{fig:graphEx}. As discussed later, the PCPM abstraction is built to easily take advantage of more sophisticated partitioning schemes and deliver further performance improvements~(the trade-off is time complexity of partitioning versus performance gains). As we show in the results section, even the simple partitioning approach described above delivers significant performance gains over state-of-the-art methods.  

Each partition is also allocated a contiguous memory space called \textit{bin} to store updates~($update\_bins$) and corresponding list of destination nodes~($destID\_bins$) in the incoming messages. Since each thread in PCPM scatters or gathers only one partition at a time, the random accesses to vertex values or partial sums are limited to address range equal to the partition size. 
%If the partition size is kept smaller than the cache, random accesses to DRAM are avoided. 
This improves temporal locality in access pattern and in turn, overall cache performance of the algorithm.
% At the same time, the amount of communication using the PNG layout and the number of address jumps made during the scatter phase is $\mathcal{O}(k^2)$~(as shown later in section~\ref{sec:analyticEval}).

Before beginning PageRank computation, each partition calculates the offsets~(address in bins where it must start writing from) into all $update\_bins$ and $destID\_bins$. Our scattering strategy dictates that the partitions write to bins in the order of their IDs. Therefore, the offset for a partition $P_i$ into any given bin is the sum of the number of values that all partitions with ID $< i$ are writing into that bin. For instance, in fig.~\ref{fig:partitioning}, the offset of partition $P_2$ into  $update\_bins[0]$ is $0$~(since partitions $P_0$ and $P_1$ do not write to bin $0$). Similarly, its offset into $update\_bins[1]$ and  $update\_bins[2]$ is $1$~(since $P_1$ writes one update to bin $1$ and $P_0$  writes one update to bin $2$). Offset computation provides each partition fixed and disjoint locations to write messages. This allows PCPM to parallelize partition processing without the need of locks or atomics. 

Note that since the destination node IDs written in the first iteration remain unchanged over the course of algorithm, they are written only once and reused in subsequent iterations. The reuse of destination node IDs along with the specific system optimizations discussed in section~\ref{sec:RR} and ~\ref{sec:PNG} enables PCPM to traverse only a fraction of the graph during scatter phase.
%The reuse of destination node IDs along with Partition-Centric update propagation~(section~\ref{sec:RR}) and PNG layout~(section~\ref{sec:PNG}) enables PCPM to traverse only a fraction of  graph in scatter phase. 
This dramatically reduces the number of DRAM accesses and \textit{gets rid of the inherent sub-optimality of GAS model}.
%\vspace{-2mm}
\subsection{Partition-Centric Update Propagation}\label{sec:RR}
\begin{figure}[htbp]
%\vspace{-1mm}
	\centering
	\begin{subfigure}{0.8\linewidth}
		\includegraphics[width=\linewidth]{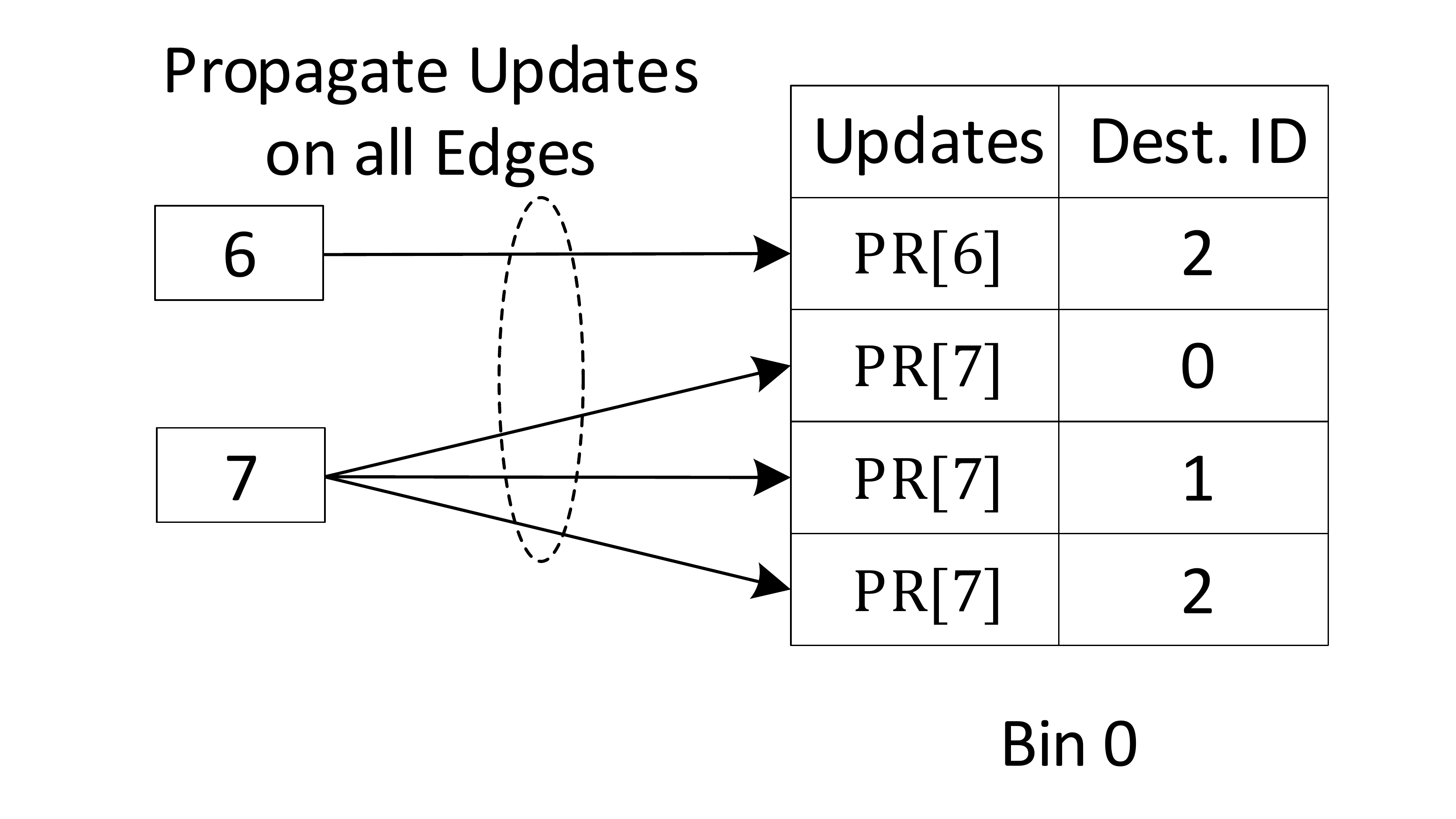}
		%\vspace{-2mm}
		\caption{Scatter in Vertex-centric GAS}
	\end{subfigure}
	%\vspace{1mm}
	\begin{subfigure}{0.8\linewidth}
		\includegraphics[width=\linewidth]{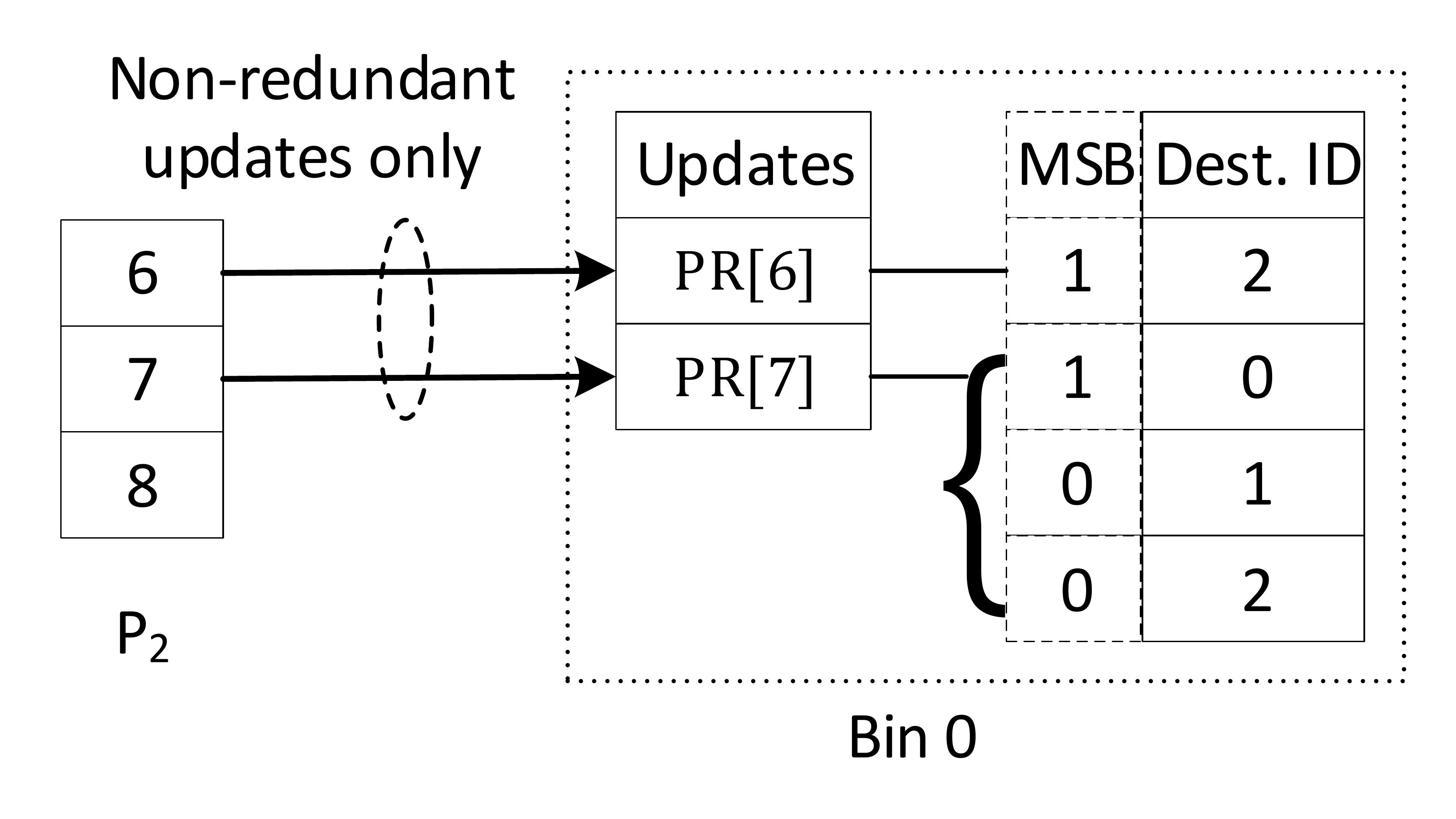}%\vspace{-2mm}
		\caption{Scatter in PCPM}
		\label{fig:pcpmScatter}
	\end{subfigure}%\vspace{1mm}
	\caption{PCPM decouples $update\_bins$ and $destID\_bins$ to avoid redundant update value propagation}
	\label{fig:scatterComp}
%\vspace{-3mm}
\end{figure}

The unique abstraction of PCPM naturally leads to transmitting a single update from a node to a neighboring partition. In other words, even if a node has multiple neighbors in a partition, it inserts only one update value in the corresponding $update\_bins$ during scatter phase~(algorithm~\ref{alg:PRPCPM}). Fig.~\ref{fig:scatterComp} illustrates the difference between Partition-Centric and Vertex-centric scatter for the example graph shown in fig.~\ref{fig:graphEx}. 

PCPM manipulates the Most Significant Bit~(MSB) of destination node IDs to indicate the range of nodes in a partition that use the same update value. In the $destID\_bins$, it consecutively writes IDs of all nodes in the neighborhood of same source vertex and sets the MSB of first ID in this range to $1$ for demarcation~(fig.~\ref{fig:pcpmScatter}). Since MSB is reserved for this functionality, PCPM supports graphs with upto 2 billion nodes instead of 4 billion for 4~Byte node IDs. However, to the best of our knowledge, this is enough to process most of the large publicly available datasets. 
%This is why, our partitioning approach allocates separate $update\_bins$ and $destID\_bins$ as shown in fig.~\ref{fig:graphBins}. 
%Destination ID lists are propagated only once and hence, not shown in fig.~\ref{fig:scatterComp}.

The gather phase starts only after all partitions are processed in the scatter phase. PCPM gather function sequentially reads updates and node IDs from the bins of the partition being processed. When gathering partition $P_i$, an update value $PR[v]$ should be applied to all out-neighbors of $v$ that lie in $P_i$. This is done by checking the MSB of node IDs to determine whether to apply the previously read update or to read the next update, as shown in algorithm~\ref{alg:PRPCPM}. The MSB is then masked to generate the true ID of destination node whose partial sum is updated.
% as shown in algorithm~\ref{alg:PRPCPM}. 

Algorithm~\ref{alg:PRPCPM} describes PCPM based PageRank computation using a row-wise partitioned CSR format for adjacency matrix $A$. Note that PCPM only writes updates for \textit{some} edges in a node's adjacency list, specifically the first outgoing edge to a partition. The remaining edges to that partition are \textit{unused}. Since CSR stores adjacencies of a node contiguously, the set of first edges to neighboring partitions is interleaved with other edges. Therefore, we have to scan all outgoing edges of each vertex during scatter phase to access this set, which decreases efficiency. Moreover, the algorithm can potentially switch bins for each update insertion, leading to random writes to DRAM. Finally, the manipulation of MSB in node indices introduces additional data dependent branches which hurts the performance. Clearly, CSR adjacency matrix is not an efficient data layout for graph processing using PCPM. In the next section, we propose a PCPM-specific data layout.
%\vspace{-1mm}
\begin{algorithm}
	\caption{PageRank iteration in PCPM using CSR format. Writing of $destID\_bins$ is not shown here.}
	\label{alg:PRPCPM}
	\begin{algorithmic}[1]
		\Statex{$q\rightarrow$ partition size, $P\rightarrow$ set of partitions}
%		\Function{Scatter}{$G(V,E), P, update\_bins$}
		\ForAll {$p \in P$} \Comment{\textbf{Scatter}}
		\ForAll {$v\in p$}
		\State{$prev\_bin \gets \infty$}
		\ForAll {$u\in N_o(v)$}
		\If {$\floor{u/q}\neq prev\_bin$}
		\State{\textbf{insert} $PR[v]$ in $update\_bins[\floor{u/q}]$}
		\State{$prev\_bin \gets \floor{u/q}$}
		\EndIf
		\EndFor
		\EndFor
		\EndFor
%		\EndFunction
%		\Function{Gather}{$G(V,E), P, update\_bins, destID\_bins$}
		\State{$PR[:] \gets 0$} 
		\ForAll {partitions $p\in P$}\Comment{\textbf{Gather}}
		\While{$destID\_bins[p]\neq\emptyset$}
		\State{\textbf{pop} $id$  from  $destID\_bins[p]$}
		\If{$MSB(id)\neq0$}
		\State{\textbf{pop} $update$  from  $update\_bins[p]$}
		\EndIf
		\State{$PR[id\ \&\ bitmask] \mathrel{+{=}} update$}
		\EndWhile
		\EndFor	
		\ForAll {$v\in V$} \Comment{\textbf{Apply}}
		\State {$PR[v] \gets \frac{(1-d)/\abs{V}\ +\ d\times PR[v]}{\abs{N_o(v)}}$}
		\EndFor
%		\EndFunction
	\end{algorithmic}
	%\vspace{-1mm}	
\end{algorithm} 
%\vspace{-3mm}
\subsection{Data Layout Optimization}\label{sec:PNG}
%\vspace{-0.5mm}
In this subsection, we describe a new bipartite Partition-Node Graph~(PNG) data layout that brings out the true Partition-Centric nature of PCPM. During the scatter phase, PNG prevents unused edge reads and ensures that all updates to a bin are streamed together before switching to another bin. 

We exploit the fact that once $destID\_bins$ are written, the only required information in PCPM is the connectivity between nodes and partitions. Therefore, edges going from a source to all destination nodes in a single partition can be \textit{compressed} into one edge whose new destination is the corresponding partition number. This gives rise to a bipartite graph $G'$ with disjoint vertex sets $V$ and $P$~(where $P=\{P_0, \ldots, P_{k-1}\}$ represents the set of partitions in the original graph), and a set of directed edges $E'$ going from $V$ to $P$. Such a transformation has the following effects:
%\vspace{-0.2mm}
\begin{enumerate}[leftmargin=*]\itemsep-0.2mm
\item \textit{Eff$_1$}$\rightarrow$ the unused edges in original graph are removed
\item \textit{Eff$_2$}$\rightarrow$ the range of destination IDs reduces from $\abs{V}$ to $\abs{P}$. 
\end{enumerate}%\vspace{-0.2mm}
The advantages of \textit{Eff$_1$} are obvious but those of \textit{Eff$_2$} will become clear when we discuss the storage format and construction of PNG.

\begin{figure}
	\centering
	\includegraphics[width=0.9\linewidth]{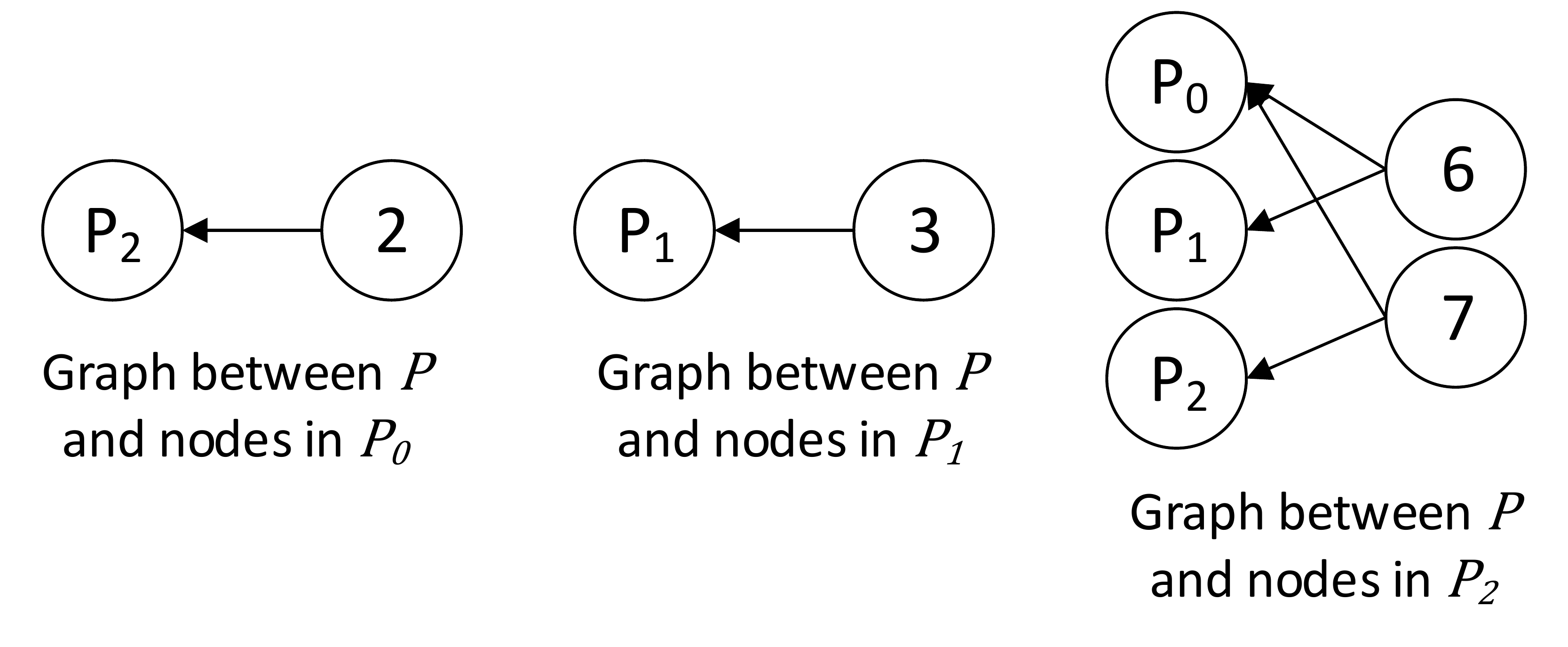}
	\caption{Partition-wise construction of PNG $G'(P,V,E')$ for graph $G(V,E)$~(fig.~\ref{fig:graphEx}). $\abs{E'}$ is much smaller than $\abs{E}$.}
	\label{fig:PNG}
	%\vspace{-4mm}
\end{figure}

The compression step reduces memory traffic by eliminating unused edge traversal. However note that scatters to a bin from source vertices in a partition are still interleaved with scatters to other bins. This can lead to random DRAM accesses during the scatter phase processing of a (source) partition. We resolve this problem by {\it transposing} the adjacency matrix of bipartite graph $G'$. The rows of the transposed matrix represent edges grouped by destination partitions which enables streaming updates to one bin at a time. This advantage comes at the cost of random accesses to source node values during the scatter phase. To prevent these random accesses from going to DRAM, we construct PNG on a per-partition basis i.e. we create a separate bipartite graph  for each partition $P_i$ with edges between $P$ and the nodes in $P_i$~(fig.~\ref{fig:PNG}). By carefully choosing $q$ to make partitions \textit{cacheable}, we ensure that all requests to source nodes are served by the cache resulting in {\it zero random DRAM accesses}. 
%The \textit{compression} step reduces memory traffic but each update insertion can still switch bins leading to random memory accesses. To tackle this problem, we \textit{transpose} the adjacency matrix of the bipartite graph. The resulting matrix stores edges grouped by destination partitions which enables streaming updates to one bin at a time. This advantage comes at the cost of random accesses to source node values during scatter phase. To prevent these random accesses from going to DRAM, we construct PNG on a per-partition basis i.e. every partition creates a separate bipartite graph between $P$ and the nodes in that partition~(fig.~\ref{fig:PNG}). Thus, for a \textit{cacheable} partition size, requests to source nodes are served by the cache resulting in \textit{zero random DRAM accesses}.

\textit{Eff$_2$} is crucial for \textit{transposition} of bipartite graphs in all partitions. The number of offsets required to store a transposed matrix in CSR format is equal to the range of destination node IDs. By reducing this range, \textit{Eff$_2$} reduces the storage requirement for offsets of each matrix from $O(\abs{V})$ to $O(\abs{P})$. Since there are $\abs{P}$ partitions, each having one bipartite graph, the total storage requirement for edge offsets in PNG is $O(\abs{P}^2)$ instead of $O(\abs{V}\times\abs{P})$.

Although PNG construction looks like a $2$-step approach, we actually merge \textit{compression} and \textit{transposition} into a single step. We first scan the outgoing edges of all nodes in a partition and individually compute the in-degree of all the destination partitions while discarding unused edges. A prefix sum of these degrees is carried out to compute the offsets array for CSR matrix. The same offsets can also be used to allocate disjoint writing locations into the bins of destination partitions. In the next scan, the edge array in CSR is filled with source node IDs completing both \textit{compression} and \textit{transposition}. PNG construction can be easily parallelized over all partitions to accelerate the pre-processing effectively.

Algorithm~\ref{alg:PCPM_PNG} shows the pseudocode for PCPM scatter phase using PNG layout. Unlike algorithm~\ref{alg:PRPCPM}, the scatter function in algorithm~\ref{alg:PCPM_PNG} does not contain data dependent branches to check and discard unused edges. Using PNG provides drastic performance gains in PCPM scatter phase with little pre-processing overhead.
%\vspace{-1mm}
\begin{algorithm}
	\caption{PCPM scatter phase using PNG layout}
	\label{alg:PCPM_PNG}
	\begin{algorithmic}[1]
		\Statex{$G'(P,V,E')\rightarrow$ PNG, $N_i^p(p')\rightarrow$ in-neighbors of partition $p'$ in bipartite graph of partition $p$}
%		\Function{Scatter}{$PNG(V,P,E'), update\_bins$}
			\ForAll {$p \in P$} \Comment{\textbf{Scatter}}
				\ForAll {$p'\in P$}
					\ForAll {$u\in N_i^p(p')$}
						\State{\textbf{insert} $PR[u]$ into $update\_bins[p']$}
					\EndFor
				\EndFor
			\EndFor
%		\EndFunction
	\end{algorithmic}
%\vspace{-1mm}	
\end{algorithm} %\vspace{-4mm}
\subsection{Branch Avoidance}\label{sec:BA}%\vspace{-1mm}
Data dependent branches have been shown to have significant impact on performance of graph algorithms~\cite{branchAvoiding} and PNG removes such branches in PCPM scatter phase. In this subsection, we propose a branch avoidance mechanism for the PCPM gather phase. Branch avoidance enhances the sustained memory bandwidth but does not impact the amount of DRAM communication. 

Note that the \textbf{pop} operations shown in algorithm~\ref{alg:PRPCPM} are implemented using pointers that increment after reading an entry from the respective bin. Let $destID\_ptr$ and $update\_ptr$ be the pointers to $destID\_bins[p]$ and $update\_bins[p]$, respectively. Note that the $destID\_ptr$ is incremented in every iteration whereas the $update\_ptr$ is only incremented if $MSB[id]\neq0$.

To implement the branch avoiding gather function, instead of using a condition check over $MSB(id)$, we add it directly to $update\_ptr$. When $MSB(id)$ is $0$, the pointer is not incremented and the same update value is read from cache in the next iteration; when $MSB(id)$ is $1$, the pointer is incremented executing the \textbf{pop} operation on $update\_bins[p]$. The modified pseudocode for gather phase is shown in algorithm~\ref{alg:BAPCPM}.
%\vspace{-2mm}
\begin{algorithm}
	\caption{Branch Avoiding gather function in PCPM}
	\label{alg:BAPCPM}
	\begin{algorithmic}[1]
%		\Function{Gather}{$V, P, update\_bins, destID\_bins$}
			\State{$PR[:]=0$}
			\ForAll {partitions $p\in P$} \Comment{\textbf{Gather}}
				\State{$\{destID\_ptr,\ update\_ptr\} \gets 0$}
				\While{$destID\_ptr\ <\ size(destID\_bins[p])$}
					\State{$id \gets destID\_bins[p][destID\_ptr\mathrel{+{+}}]$}
					\State{$update\_ptr \mathrel{+{=}} MSB(id)$}
					\State{$id \gets id\ \&\ bitmask$}
					\State{$PR[id] \mathrel{+{=}} update\_bins[p][update\_ptr]$}
				\EndWhile
			\EndFor	
%			\ForAll {$v\in V$} \Comment{\textbf{Apply}}
%				\State {$PR[v] \gets \frac{(1-d)/\abs{V}\ +\ d\times PR[v]}{\abs{N_o(v)}}$}
%			\EndFor
%		\EndFunction
	\end{algorithmic}
%\vspace{-1mm}	
\end{algorithm}
%\vspace{-1mm}
\subsection{Weighted Graphs and SpMV}\label{sec:ext}
PCPM can be easily extended for computation on weighted graphs by storing the edge weights along with destination IDs in $destID\_bins$. These weights can be read in the gather phase and applied to the source node value before updating the destination node. PCPM can also be extended to generic SpMV with non-square matrices by partitioning the rows and columns separately. In this case, the outermost loops in scatter phase~(algorithm~\ref{alg:PCPM_PNG}) and gather phase~(algorithm~\ref{alg:BAPCPM}) will iterate over row partitions and column partitions of $A$, respectively.
%\vspace{-2mm}
\subsection{Comparison with Vertex-centric GAS}\label{sec:algoComp}
The Binning with Vertex-centric GAS~(BVGAS) method allocates multiple bins to store incoming messages~($(update, destID)$ pairs). If bin width is $q$, then all messages destined to $v\in [i*q, (i+1)*q)$ are written in bin $i$. The scatter phase traverses the graph in a Vertex-centric fashion and inserts the messages in respective bins of the destination vertices. Number of bins is kept small to allow insertion points for all bins to fit in cache, providing good spatial locality. The gather phase processes one bin at a time as shown in algorithm~\ref{alg:PRBVGAS}, and thus, enjoys good temporal locality if bin width is small. 
%\vspace{-1mm}
\begin{algorithm}
		\caption{PageRank Iteration using BVGAS}
		\label{alg:PRBVGAS}
		\begin{algorithmic}[1]
			\Statex{$q\rightarrow$ bin width, $B\rightarrow$ no. of bins}
			\For {$v\in V$} \Comment{\textbf{Scatter}}
			\State{$PR[v] = PR[v]/\abs{N_o(v)}$}
			\ForAll {$u\in N_o(v)$}
			\State{\textbf{insert} $(PR[v], u)$ into $bins[\floor{u/q}]$}
			\EndFor
			\EndFor
			\State{$PR[:]=0$}
			\For {$b = 0$  to  $B-1$} \Comment{\textbf{Gather}}
			\ForAll {$(update, dest)$ in $bins[b]$}
			\State {$PR[dest] = PR[dest] + update$}
			\EndFor	
			\EndFor
			\ForAll {$v\in V$} \Comment{\textbf{Apply}}
			\State {$PR[v] = \frac{(1-d)}{\abs{V}}\ +\ d\times PR[v] $}
			\EndFor
		\end{algorithmic}
	%\vspace{-1mm}
\end{algorithm}

Unlike algorithm~\ref{alg:PRBVGAS}, in our BVGAS implementation, we write the destination IDs only in the first iteration. We also use small cached buffers to store updates before writing to DRAM. This ensures full cache line utilization and reduces communication during scatter phase~\cite{propagationBlocking}.

Irrespective of all the locality advantages and optimizations, BVGAS inherently suffers from redundant reads and writes of a vertex value on all of its outgoing edges. This redundancy manifests itself in the form of BVGAS' inability to utilize \textit{high locality} in graphs with optimized node labeling. PCPM on the other hand, uses graph locality to reduce the fraction of graph traversed in scatter phase. Unlike PCPM, the Vertex-centric traversal in BVGAS can also insert consecutive updates into different bins. This leads to random DRAM accesses and poor bandwidth utilization. We provide a quantitative analysis of these differences in the next section.
%\vspace{-0.5mm}
\section{Analytical Evaluation}\label{sec:analyticEval}
We derive performance models to compare PCPM against conventional Pull Direction PageRank~(PDPR) and BVGAS. Our models provide valuable insights into the behavior of different methodologies with respect to varying graph structure and locality. Table~\ref{table:params} defines the parameters used in the analysis. We use a synthetic kronecker graph~\cite{graph500} of scale 25~\textit{(kron)} as an example for illustration purposes.
\begin{table}[]
	\centering
	\caption{List of model parameters}
	\label{table:params}
	\resizebox{\linewidth}{!}{%	
	\begin{tabular}{cccc}
		\cline{1-2} \cline{3-4}
		\multicolumn{2}{|c||}{\textbf{Original Graph $G(V,E)$}}                                                                                   & \multicolumn{2}{c|}{\textbf{PNG layout $G'(P, V, E')$}}                            \\ \cline{1-2} \cline{3-4} 
		\multicolumn{1}{|c|}{$n$}   & \multicolumn{1}{c||}{no. of vertices~(\abs{V})}                                                           & \multicolumn{1}{c|}{$k$} & \multicolumn{1}{c|}{no. of partitions~(\abs{P})}          \\ \cline{1-2} \cline{3-4} 
		\multicolumn{1}{|c|}{$m$}   & \multicolumn{1}{c||}{no. of edges~(\abs{E})}                                                             &  \multicolumn{1}{c|}{$r$} & \multicolumn{1}{c|}{compression ratio~(\nicefrac{\abs{E}}{\abs{E'}})} \\ \hline \hline
		\multicolumn{2}{|c||}{\textbf{Architecture}}                                                                                              & \multicolumn{2}{c|}{\textbf{Software}}                                             \\ \cline{1-2} \cline{3-4} 
		\multicolumn{1}{|c|}{$c_{mr}$} & \multicolumn{1}{c||}{\begin{tabular}[c]{@{}c@{}}cache miss ratio for source\\    value reads in PDPR\end{tabular}} &  \multicolumn{1}{c|}{$d_v$} & \multicolumn{1}{c|}{sizeof~(updates/PageRank value)}          \\ \cline{1-2} \cline{3-4} 
		\multicolumn{1}{|c|}{$l$}  & \multicolumn{1}{c||}{sizeof~(cache line)}                                                                     & \multicolumn{1}{c|}{$d_i$} & \multicolumn{1}{c|}{sizeof~(node or edge index)}          \\ \cline{1-2} \cline{3-4} 
	\end{tabular}%
}
\end{table}
%\vspace{-1mm}\subsection{DRAM Communication}\label{sec:commModel}
 We analyze the amount of data exchanged with main memory per iteration of PageRank. We assume that data is accessed in quantum of one cache line and BVGAS exhibits full cache line utilization. Since destination indices are written only in the first iteration for PCPM and BVGAS, they are not accounted for in this model. 

\noindent\textbf{PDPR:} The pull technique scans all edges in the graph once~(algorithm~\ref{alg:PRPull}). For a CSR format, this requires reading $n$ edge offsets and $m$ source node indices. PDPR also reads $m$ source node values that incur cache misses generating $mc_{mr}l$~Bytes of DRAM traffic. Outputting new PageRank values generates $nd_v$~Bytes of writes to DRAM. The total communication volume for PDPR is:
\begin{equation}\label{eq:PDPRComm}
	PDPR_{comm} = m(d_i + c_{mr}l) + n(d_i + d_v) %\approxeq m(d_i + c_{mr}l + d_v),
\end{equation}
\noindent\textbf{BVGAS:} The scatter phase~(algorithm~\ref{alg:PRBVGAS}) scans the graph and writes updates on all outgoing edges of the source node, thus communicating $(n+m)d_i + (n+m)d_v$~Bytes. The gather phase loads updates and destination node IDs on all the edges generating $m(d_i+d_v)$~Bytes of read traffic. At the end of gather phase, $nd_v$~Bytes of new PageRank values are written in the main memory. Total communication volume for BVGAS is therefore, given by:
\begin{equation}\label{eq:BVGASComm}
	BVGAS_{comm} = 2m(d_i+d_v)+n(d_i+2d_v) %\approxeq 2m(d_i + d_v)
\end{equation}
\noindent\textbf{PCPM with PNG:} Number of edge offsets in bipartite graph of each partition is $k$. Thus, in the scatter phase~(algorithm~\ref{alg:PCPM_PNG}), a scan of PNG reads $(k\times k + \nicefrac{m}{r})d_i$~Bytes. The scatter phase further reads $n$ PageRank values and writes updates on $\nicefrac{m}{r}$ edges. The gather phase~(algorithm~\ref{alg:BAPCPM}) reads $m$ destination IDs and $\nicefrac{m}{r}$ updates followed by $n$ new PageRank value writes. Net communication volume in PCPM is given by:
\begin{equation}\label{eq:PCPMComm}
	PCPM_{comm} = m\left(d_i\left(1+\frac{1}{r}\right) + \frac{2d_v}{r}\right) + k^2d_i + 2nd_v 
\end{equation}
%\begin{itemize}
%	\item Glossary of model and parameters
%	\item communication for pull direction and GAS vertex centric
%	\item reduction in net communication with redundancy reduction (reduces writes) and bin graph(reduces reads and writes both)
%	\item reduction in the number of jumps in memory accesses with bin graph transposing
%	\item reduction in the number of branch misses during gather with branch avoidance
%\end{itemize}

\noindent\textbf{Comparison:} Performance of pull technique depends heavily on $c_{mr}$. In the worst case, all accesses are cache misses i.e. $c_{mr}=1$ and in best case, only cold misses are encountered to load the PageRank values in cache i.e. $c_{mr}=\nicefrac{nd_v}{ml}$. Assuming $k^2 \ll n \ll m$, we get $PDPR_{comm} \in [md_i, m(d_i+l)]$. On the other hand, communication for BVGAS stays constant. With $\theta(m)$ additional loads and stores, $BVGAS_{comm}$ can never reach the lower bound of $PDPR_{comm}$. Comparatively, $PCPM_{comm}$ achieves optimality when for every vertex, all outgoing edges can be compressed into a single edge i.e. $r=\nicefrac{m}{n}$. In the worst case when $r=1$, PCPM is still as good as BVGAS and we get $PCPM_{comm}\in[md_i, m(2d_i+2d_v)]$. Unlike BVGAS, $PCPM_{comm}$ achieves the same lower bound as $PDPR_{comm}$.

Analyzing equations~\ref{eq:PDPRComm} and~\ref{eq:BVGASComm}, we see that BVGAS is profitable compared to PDPR when:
\begin{equation}\label{eq:BVGASComp}
c_{mr} > \frac{d_i+2d_v}{l}
\end{equation}
In comparison, PCPM offers a more relaxed constraint on $c_{mr}$~(by a factor of \nicefrac{1}{r}) becoming advantageous when:
\begin{equation}\label{eq:PCPMComp}
c_{mr} > \frac{d_i+2d_v}{rl}
\end{equation}

\begin{figure}
	\includegraphics[width=0.9\linewidth]{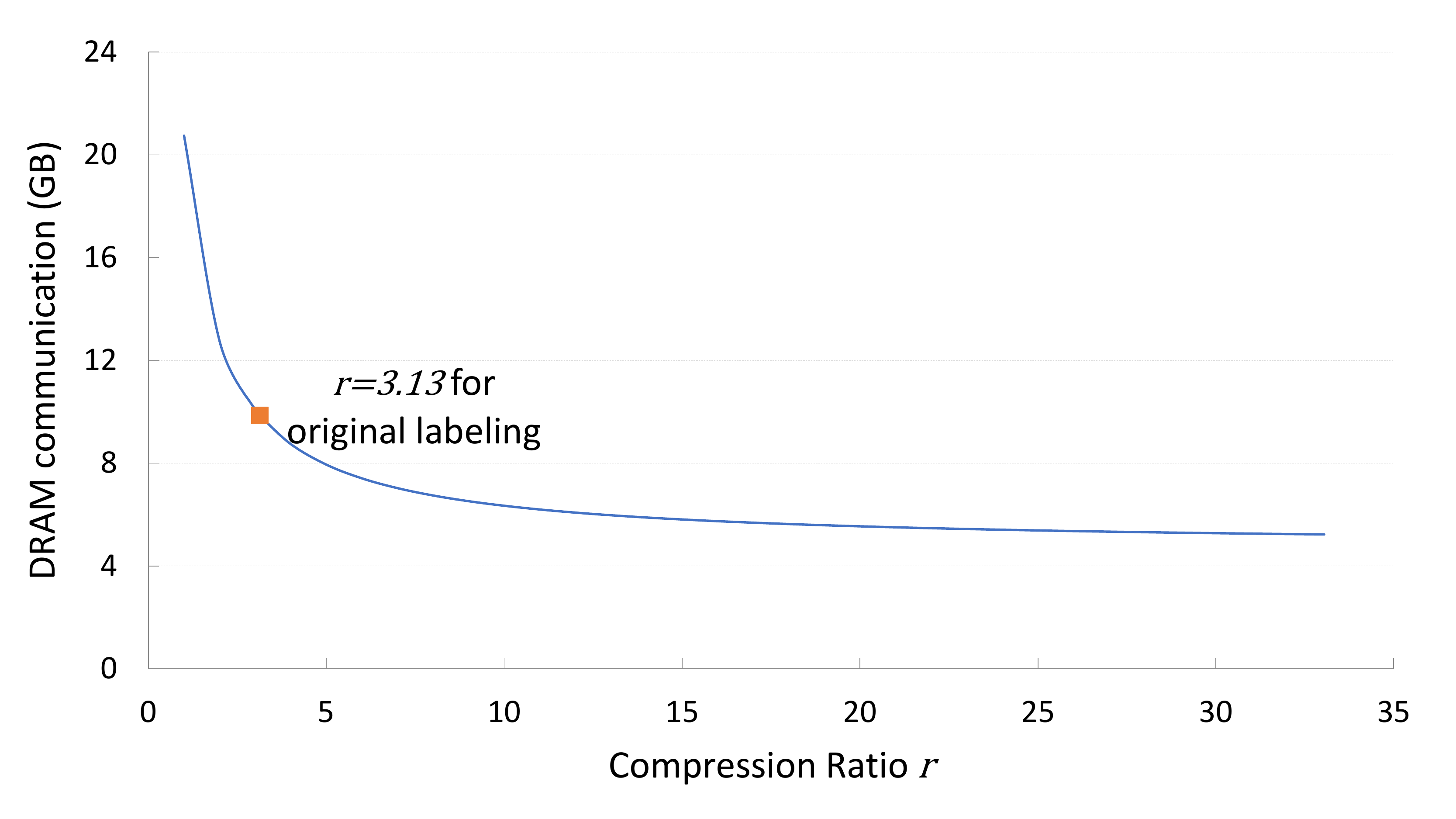}
	\caption{Predicted DRAM traffic for \textit{kron} graph with $n=33.5$~M, $m=1070$~M, $k=512$ and $d_i=d_v=4$~Bytes.}
	\label{fig:commModel}
\end{figure}

The RHS in eq.~\ref{eq:BVGASComp} is constant indicating that BVGAS is advantageous for low locality graphs. With optimized node ordering, we can reduce $c_{mr}$ and outperform BVGAS. On the contrary, $r\in[1,\nicefrac{m}{n}]$ in the RHS of eq.~\ref{eq:PCPMComp} is a function of locality. With an optimized node labeling, $r$ also increases and enhances the performance of PCPM. Fig.~\ref{fig:commModel} shows the effect of $r$ on predicted DRAM communication for the \textit{kron} graph. Obtaining an optimal nodel labeling that makes $r=\nicefrac{m}{n}$ might be very difficult or even impossible for some graphs. However, as can be observed from fig.~\ref{fig:commModel}, DRAM traffic decreases rapidly for $r\leq5$ and converges slowly for $r>5$. Therefore, a node reordering that can achieve $r\approx5$ is good enough to optimize communication volume in PCPM.
\subsection{Random Memory Accesses}
We define a random access as a non-sequential jump in the address of memory location being read from or written to DRAM. Random accesses can incur latency penalties and negatively impact the sustained memory bandwidth. In this subsection, we model the amount of random accesses performed by different methodologies in a single PageRank iteration.

\noindent\textbf{PDPR:} %Computing random accesses for pull technique is fairly straightforward. 
Reading edge offsets and source node IDs in pull technique is completely sequential because of the CSR format. However, all accesses to source node PageRank values served by DRAM contribute to potential random accesses resulting in:
\begin{equation}
	PDPR_{ra} = O(mc_{mr})
\end{equation}
\noindent\textbf{BVGAS:} In scatter phase of algorithm~\ref{alg:PRBVGAS}, updates can potentially be inserted at random memory locations. Assuming full cache line utilization for BVGAS, for every $l$~Bytes written, there is at most $1$ random DRAM access. In gather phase, all DRAM accesses are sequential if we assume that bin width is smaller than the cache. Total random accesses for BVGAS are then given by:
\begin{equation}
	BVGAS_{ra} = O\left(\frac{md_v}{l}\right)
\end{equation}
\noindent\textbf{PCPM:} With the PNG layout~(algorithm~\ref{alg:PCPM_PNG}), there are at most $k$ bin switches when scattering updates from a partition. Since there are $k$ such partitions, total number of random accesses in PCPM is bound by:
\begin{equation}
	PCPM_{ra} = O(k*k) = O(k^2)
\end{equation}
\noindent\textbf{Comparison:} BVGAS exhibits less random accesses than PDPR. However, $PCPM_{ra}$ is much smaller than both $BVGAS_{ra}$ and $PDPR_{ra}$. For instance, in the \textit{kron} dataset with $d_v=4$~Bytes, $l=64$~Bytes and $k=512$, $BVGAS_{ra} \approx 66.9$~M whereas $PCPM_{ra}\approx 0.26$~M. 

Although it is not indicated in algorithm~\ref{alg:PRBVGAS}, the number of data dependent unpredictable branches in cache bypassing BVGAS implementation is also $O(m)$. For every update insertion, the BVGAS scatter function has to check if the corresponding cached buffer is full~(section~\ref{sec:algoComp}). In contrast, the number of branch mispredictions for PCPM~(using branch avoidance) is $O(k^2)$ with $1$ misprediction for every destination partition~($p'$) switch in algorithm~\ref{alg:PCPM_PNG}. The derivations are similar to random access model and for the sake of brevity, we do not provide a detailed deduction. 
%%\vspace{-1mm}
\section{Experimental Evaluation}\label{sec:expEval}%%\vspace{-0.5mm}
\subsection{Experimental Setup and Datasets}
We conduct experiments on a dual-socket Ivy Bridge server equipped with two 8-core Intel Xeon E5-2650 v2 processors@2.6~GHz running Ubuntu 14.04 OS. Table~\ref{table:sysChars} lists important characteristics of our machine. Memory bandwidth is measured using STREAM benchmark~\cite{stream}.
All codes are written in C++ and compiled using G++ 4.7.1 with the highest optimization -O3 flag. The memory statistics are collected using Intel Performance Counter Monitor~\cite{pcm}. All data types used for indices and PageRank values are $4$~Bytes.
%\vspace{-1mm}
\begin{table}[htbp]
	\centering
	\caption{System Characteristics}
	\label{table:sysChars}
	\resizebox{0.7\linewidth}{!}{%	
	\begin{tabular}{|c|c|c|}
		\hline
		\multirow{2}{*}{\textbf{Socket}} & no. of cores    & 8         \\ \cline{2-3} 
		& shared L3 cache & 25MB      \\ \hline
		\multirow{2}{*}{\textbf{Core}}   & L1d  cache      & 32 KB     \\ \cline{2-3} 
		& L2 cache        & 256 KB   \\ \hline
		\multirow{3}{*}{\textbf{Memory}} & size            & 128 GB    \\ \cline{2-3} 
		& Read BW         & 59.6 GB/s \\ \cline{2-3} 
		& Write BW        & 32.9 GB/s \\ \hline
	\end{tabular}
}
%\vspace{-1mm}
\end{table}

We use 6 large real world and synthetic graph datasets coming from different applications, for performance evaluation. Table~\ref{table:datasets} summarizes the size and sparsity characteristics of these graphs.
\textit{Gplus} and \textit{twitter} are follower graphs on social networks; \textit{pld}, \textit{web} and \textit{sd1} are hyperlink graphs obtained by web crawlers; and \textit{kron} is a scale 25 graph generated using Graph500 Kronecker generator. The \textit{web} is a very sparse graph but has high locality obtained by a very expensive pre-processing of node labels~\cite{webbase}. 
The \textit{kron} graph has higher edge density as compared to other datasets.
\begin{table}[htbp]
	\centering
	\caption{Graph Datasets}
	\label{table:datasets}
	\resizebox{\linewidth}{!}{%
		\begin{tabular}{|c|c|c|c|c|}
			\hline
			\textbf{Dataset} & \textbf{Description} & \textbf{\# Nodes (M)} & \textbf{\# Edges (M)} & \textbf{Degree} \\ \hline
			gplus~\cite{gplus}           & Google Plus          & 28.94                    & 462.99                & 16              \\ \hline
			pld~\cite{domain}              & Pay-Level-Domain     & 42.89                    & 623.06                & 14.53           \\ \hline
			web~\cite{webbase}              & Webbase-2001         & 118.14                   & 992.84                & 8.4             \\ \hline
			kron~\cite{graph500}           & Synthetic graph  & 33.5                     & 1047.93               & 31.28           \\ \hline
			twitter~\cite{twitter}          & Follower network     & 61.58                    & 1468.36               & 23.84           \\ \hline
			sd1~\cite{domain}              & Subdomain graph      & 94.95                    & 1937.49               & 20.4            \\ \hline
		\end{tabular}
	}
%\vspace{-1mm}
\end{table}
%\vspace{-3mm}
\subsection{Implementation Details}
We use a simple hand coded implementation of algorithm~\ref{alg:PRPull} for PDPR and parallelize it over vertices with static load balancing on the number of edges traversed. Our baseline does not incur overheads associated with similar implementations in frameworks~\cite{ligra,galois,graphmat} and hence, is faster than framework based programs~\cite{propagationBlocking}.

To parallelize BVGAS scatter phase~(algorithm~\ref{alg:PRBVGAS}), we give each thread a fixed range of nodes to scatter. Work per thread is statically balanced in terms of the number of edges processed. We also give each thread distinct memory spaces corresponding to all bins to avoid atomicity concerns in scatter phase. We use the Intel AVX non-temporal store instructions~\cite{intel} to bypass the cache while writing updates and use $128$~Bytes cache line aligned buffers to accumulate the updates for streaming stores~\cite{propagationBlocking}. BVGAS gather phase is parallelized over bins with load balanced using OpenMP dynamic scheduling. The optimal bin width is empirically determined and set to $256$~KB~($64$K nodes). As bin width is a power of $2$, we use bit shift instructions instead of integer division to compute the destination bin from node ID.

The PCPM scatter and gather phases are parallelized over partitions and load balancing in both the phases is done dynamically using OpenMP. Partition size is empirically determined and set to $256$~KB. A detailed design space exploration of PCPM is discussed in section~\ref{sec:paramDeterm}. 

All the implementations mentioned in this section execute $20$ PageRank iterations on $16$ cores. For accuracy of the collected information, we repeat these algorithms $5$ times and report the average values.
%\vspace{-0.5mm}
\subsection{Results}%\vspace{-0.5mm}
\subsubsection{Comparison with Baselines}
\begin{figure}[htbp]
	\includegraphics[width=\linewidth]{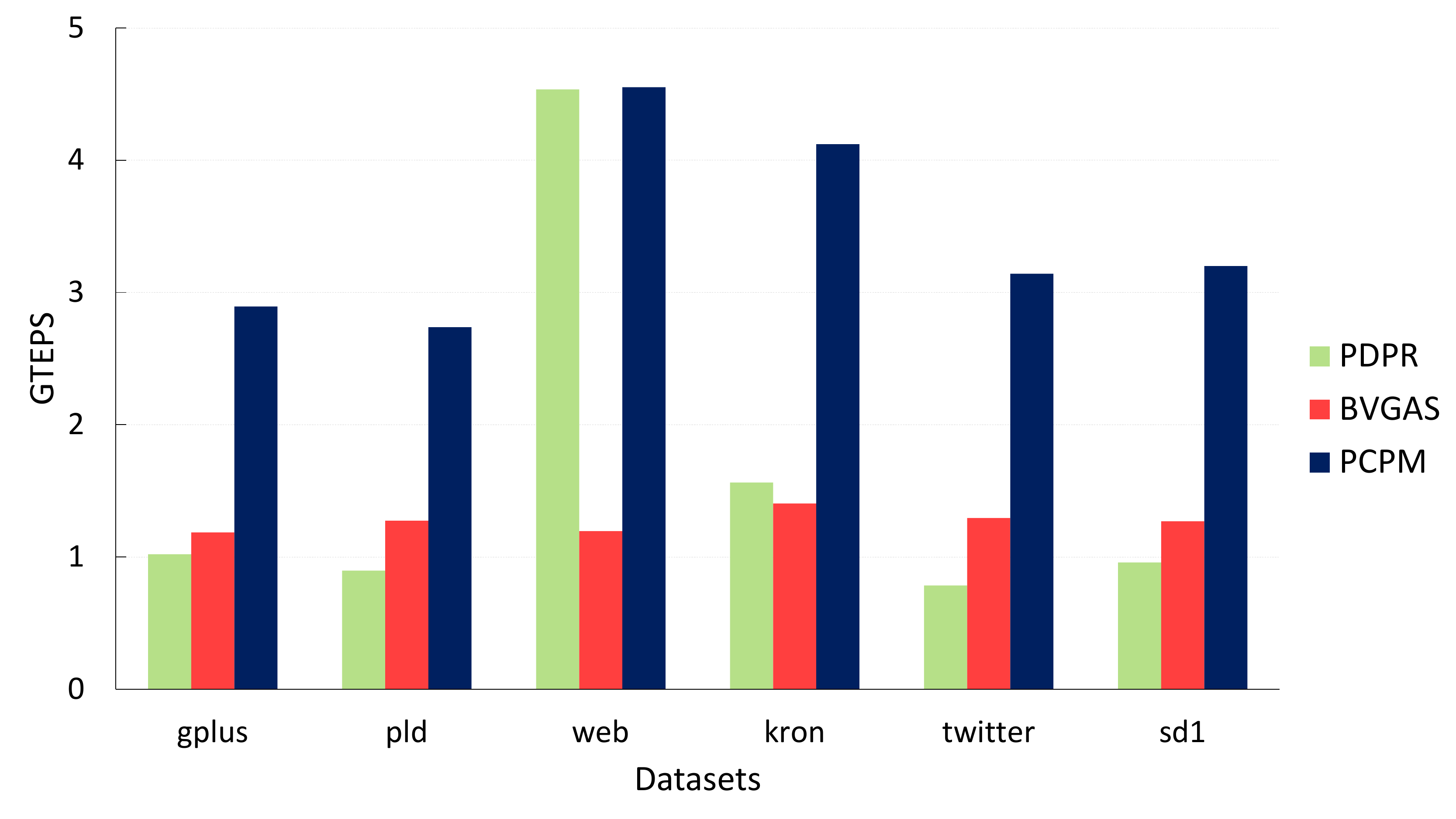}
	\caption{Performance in GTEPS. PCPM provides substantial speedup over BVGAS and PDPR.}
	\label{fig:gteps}
%	%\vspace{-5mm}
\end{figure}

%In this section we compare PCPM against the PDPR and BVGAS baselines using various criterion.
\noindent\textbf{Execution Time:} Fig.~\ref{fig:gteps} gives a comparison of the GTEPS~(computed as the ratio of giga edges in the graph to the runtime of single PageRank iteration) achieved by different implementations. We observe that PCPM is $2.1-3.8\times$ faster than the state-of-the-art BVGAS implementation and upto $4.1\times$ faster than PDPR. BVGAS achieves constant throughput irrespective of the graph structure and is able to accelerate computation on low locality graphs. However, it is worse than PDPR for high locality~(\textit{web}) and dense~(\textit{kron}) graphs. PCPM is able to outperform PDPR and BVGAS on all datasets, though the speedup on \textit{web} graph is minute because of high performance of PDPR.
Detailed results for execution time of BVGAS and PCPM during different phases of computation are given in table~\ref{table:execTime}. PCPM scatter phase benefits from a multitude of optimizations to achieve a dramatic $~5\times$ speedup over BVGAS scatter phase.
\begin{table}[htbp]
	\centering
	\caption{Execution time per iteration of PageRank for PDPR, BVGAS and PCPM}%\vspace{-1mm}
	\label{table:execTime}
	\resizebox{\linewidth}{!}{%	
\begin{tabular}{cccccccc}
	\textbf{}                     & \textbf{PDPR}                                                                & \multicolumn{3}{c}{\textbf{BVGAS}}                                                                                                                                                                                                              & \multicolumn{3}{c}{\textbf{PCPM}}                                                                                                                                                                                                               \\ \hline
	\multicolumn{1}{|c|}{Dataset} & \multicolumn{1}{c|}{\begin{tabular}[c]{@{}c@{}}Total\\ Time(s)\end{tabular}} & \multicolumn{1}{c}{\begin{tabular}[c]{@{}c@{}}Scatter \\ Time(s)\end{tabular}} & \multicolumn{1}{c}{\begin{tabular}[c]{@{}c@{}}Gather \\ Time(s)\end{tabular}} & \multicolumn{1}{c|}{\begin{tabular}[c]{@{}c@{}}Total\\ Time(s)\end{tabular}} & \multicolumn{1}{c}{\begin{tabular}[c]{@{}c@{}}Scatter\\  Time(s)\end{tabular}} & \multicolumn{1}{c}{\begin{tabular}[c]{@{}c@{}}Gather \\ Time(s)\end{tabular}} & \multicolumn{1}{c|}{\begin{tabular}[c]{@{}c@{}}Total\\ Time(s)\end{tabular}} \\ \hline
	\multicolumn{1}{|c|}{gplus}   & \multicolumn{1}{c|}{0.44}                                                    & \multicolumn{1}{c}{0.26}                                                       & \multicolumn{1}{c}{0.12}                                                      & \multicolumn{1}{c|}{0.38}                                                    & \multicolumn{1}{c}{0.06}                                                       & \multicolumn{1}{c}{0.1}                                                       & \multicolumn{1}{c|}{0.16}                                                    \\ \hline
	\multicolumn{1}{|c|}{pld}     & \multicolumn{1}{c|}{0.68}                                                    & \multicolumn{1}{c}{0.33}                                                       & \multicolumn{1}{c}{0.15}                                                      & \multicolumn{1}{c|}{0.48}                                                    & \multicolumn{1}{c}{0.09}                                                       & \multicolumn{1}{c}{0.13}                                                      & \multicolumn{1}{c|}{0.22}                                                    \\ \hline
	\multicolumn{1}{|c|}{web}     & \multicolumn{1}{c|}{0.21}                                                    & \multicolumn{1}{c}{0.58}                                                       & \multicolumn{1}{c}{0.23}                                                      & \multicolumn{1}{c|}{0.81}                                                    & \multicolumn{1}{c}{0.04}                                                       & \multicolumn{1}{c}{0.17}                                                      & \multicolumn{1}{c|}{0.21}                                                    \\ \hline
	\multicolumn{1}{|c|}{kron}    & \multicolumn{1}{c|}{0.65}                                                    & \multicolumn{1}{c}{0.5}                                                        & \multicolumn{1}{c}{0.22}                                                      & \multicolumn{1}{c|}{0.72}                                                    & \multicolumn{1}{c}{0.07}                                                       & \multicolumn{1}{c}{0.18}                                                      & \multicolumn{1}{c|}{0.25}                                                    \\ \hline
	\multicolumn{1}{|c|}{twitter} & \multicolumn{1}{c|}{1.83}                                                    & \multicolumn{1}{c}{0.79}                                                       & \multicolumn{1}{c}{0.32}                                                      & \multicolumn{1}{c|}{1.11}                                                    & \multicolumn{1}{c}{0.18}                                                       & \multicolumn{1}{c}{0.27}                                                      & \multicolumn{1}{c|}{0.45}                                                    \\ \hline
	\multicolumn{1}{|c|}{sd1}     & \multicolumn{1}{c|}{1.97}                                                    & \multicolumn{1}{c}{1.07}                                                       & \multicolumn{1}{c}{0.42}                                                      & \multicolumn{1}{c|}{1.49}                                                    & \multicolumn{1}{c}{0.24}                                                       & \multicolumn{1}{c}{0.35}                                                      & \multicolumn{1}{c|}{0.59}                                                    \\ \hline
\end{tabular}%
}%\vspace{-2mm}
\end{table}

\begin{figure}[htbp]
	\includegraphics[width=\linewidth]{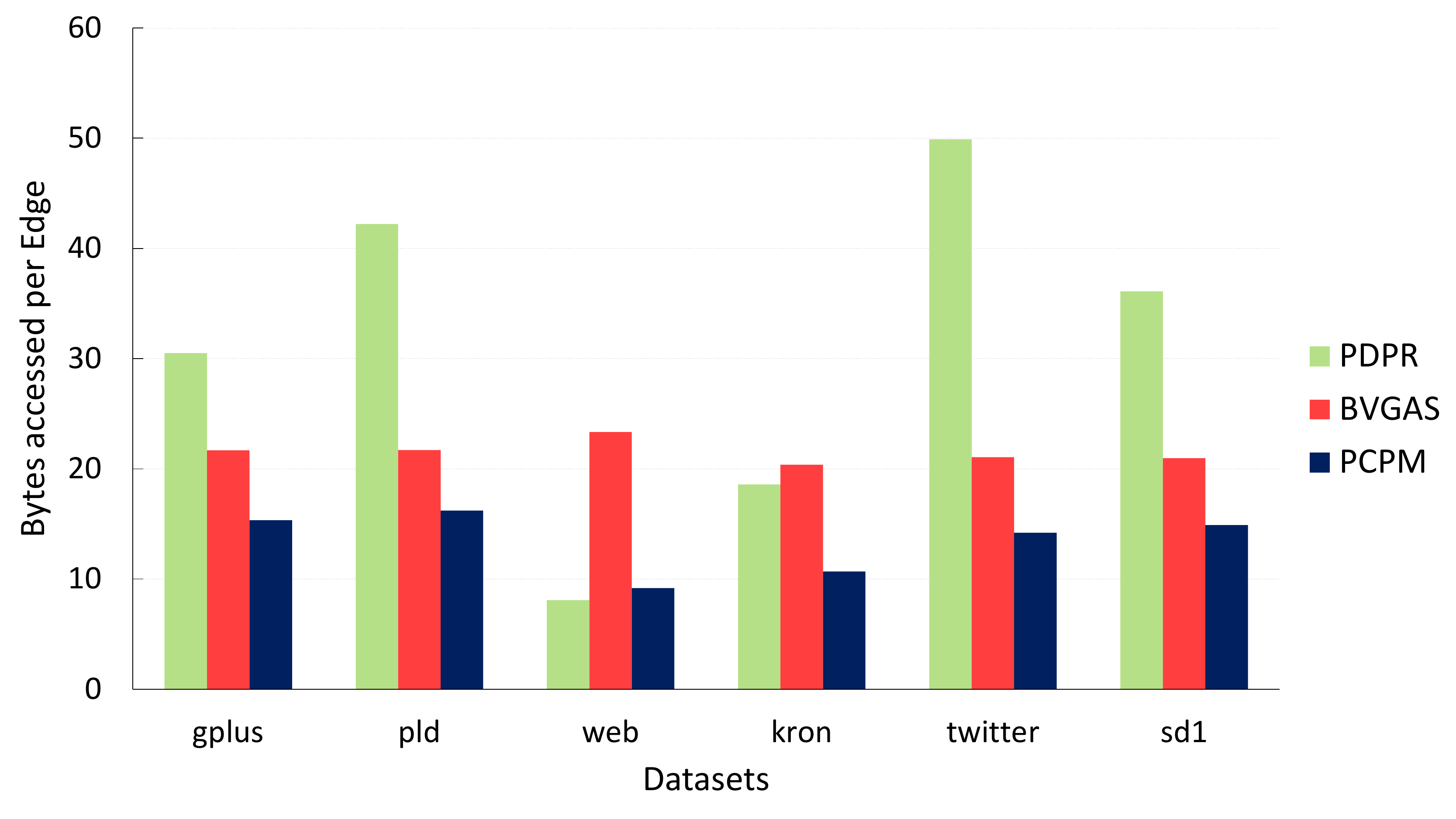}
	\caption{Main memory traffic per edge. PCPM communicates the least for all datasets except the \textit{web} graph.}
	\label{fig:bytesPerEdge}%\vspace{-2mm}
\end{figure}

\noindent\textbf{Communication and Bandwidth:} Fig.~\ref{fig:bytesPerEdge} shows the amount of data communicated with main memory normalized by the number of edges in the graph. Average communication in PCPM is $1.7\times$ and $2.2\times$ less than BVGAS and PDPR, respectively. Further, PCPM memory traffic per edge for \textit{web} and \textit{kron} is lower than other graphs because of their high compression ratio~(table~\ref{table:compRatio}). The normalized communication for BVGAS is almost constant and therefore, its utility depends on the efficiency of pull direction baseline.

\begin{figure}[htbp]
	\includegraphics[width=\linewidth]{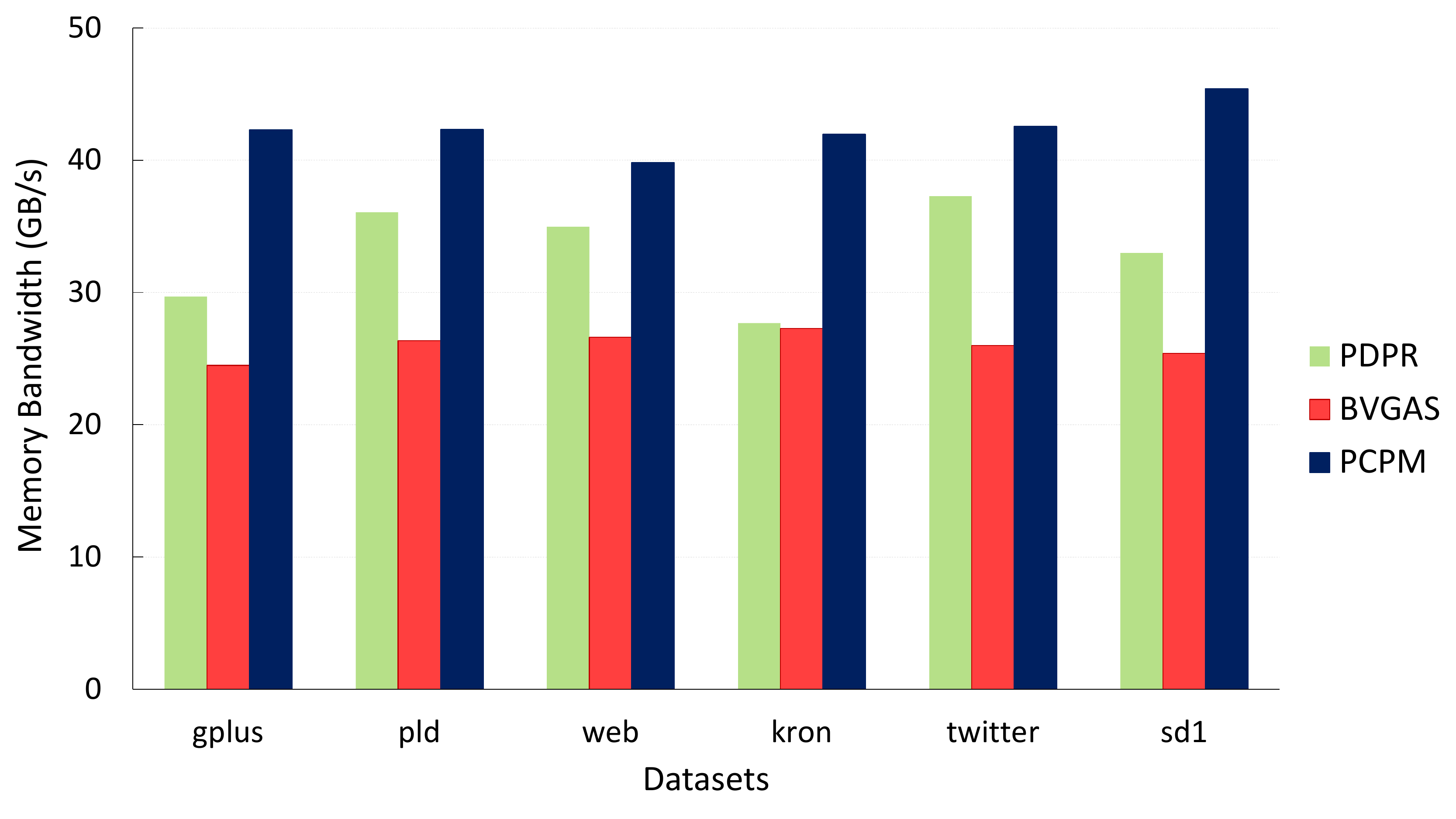}
	\caption{Sustained Memory Bandwidth for different methods. PCPM achieves highest bandwidth utilization.}
	\label{fig:memBW}%\vspace{-4mm}
\end{figure}

Note that the speedup obtained by PCPM is larger than the reduction in communication volume. This is because by avoiding random DRAM accesses and unpredictable branches, PCPM is able to efficiently utilize the available DRAM bandwidth. As shown in fig.~\ref{fig:memBW}, PCPM can sustain an average $42.4$~GB/s bandwidth compared to $33.1$~GB/s and $26$~GB/s of PDPR and BVGAS, respectively. For large graphs like \textit{sd1}, PCPM achieves $\approx77\%$ of the peak read bandwidth~(table~\ref{table:sysChars}) of our system. Although both PDPR and BVGAS suffer from random memory accesses, the former executes very few instructions and therefore, has better bandwidth utilization.  
\begin{table}[htbp]
	\centering
	\caption{Locality vs compression ratio $r$. GOrder improves locality in neighbors and increases compression}%\vspace{-1mm}
	\label{table:compRatio}
	\resizebox{\linewidth}{!}{%	
	\begin{tabular}{cccccc}
		\textbf{}                     & \textbf{}                                                                              & \multicolumn{2}{c}{\textbf{Original Labeling}}                                                                     & \multicolumn{2}{c}{\textbf{GOrder Labeling}}                                                                      \\ \hline
		\multicolumn{1}{|c|}{Dataset} & \multicolumn{1}{c|}{\begin{tabular}[c]{@{}c@{}}\#Edges in\\ Graph~(M)\end{tabular}} & \multicolumn{1}{c}{\begin{tabular}[c]{@{}c@{}}\#Edges in \\ PNG~(M)\end{tabular}} & \multicolumn{1}{c|}{$r$}  & \multicolumn{1}{c}{\begin{tabular}[c]{@{}c@{}}\#Edges in \\ PNG~(M)\end{tabular}} & \multicolumn{1}{c|}{$r$}  \\ \hline
		\multicolumn{1}{|c|}{gplus}   & \multicolumn{1}{c|}{463}                                                               & \multicolumn{1}{c}{243.8}                                                              & \multicolumn{1}{c|}{1.9}  & \multicolumn{1}{c}{157.4}                                                              & \multicolumn{1}{c|}{2.94} \\ \hline
		\multicolumn{1}{|c|}{pld}     & \multicolumn{1}{c|}{623.1}                                                             & \multicolumn{1}{c}{347.7}                                                              & \multicolumn{1}{c|}{1.79} & \multicolumn{1}{c}{166.7}                                                              & \multicolumn{1}{c|}{3.73} \\ \hline
		\multicolumn{1}{|c|}{web}     & \multicolumn{1}{c|}{992.8}                                                             & \multicolumn{1}{c}{118.1}                                                              & \multicolumn{1}{c|}{8.4}  & \multicolumn{1}{c}{126.8}                                                              & \multicolumn{1}{c|}{7.83} \\ \hline
		\multicolumn{1}{|c|}{kron}    & \multicolumn{1}{c|}{104.8}                                                             & \multicolumn{1}{c}{342.7}                                                              & \multicolumn{1}{c|}{3.06} & \multicolumn{1}{c}{169.7}                                                              & \multicolumn{1}{c|}{6.17} \\ \hline
		\multicolumn{1}{|c|}{twitter} & \multicolumn{1}{c|}{1468.4}                                                            & \multicolumn{1}{c}{722.4}                                                              & \multicolumn{1}{c|}{2.03} & \multicolumn{1}{c}{386.2}                                                              & \multicolumn{1}{c|}{3.8}  \\ \hline
		\multicolumn{1}{|c|}{sd1}     & \multicolumn{1}{c|}{1937.5}                                                            & \multicolumn{1}{c}{976.9}                                                              & \multicolumn{1}{c|}{1.98} & \multicolumn{1}{c}{366.2}                                                              & \multicolumn{1}{c|}{5.29} \\ \hline
	\end{tabular}%
}
\end{table}

\begin{figure}
	\includegraphics[width=\linewidth]{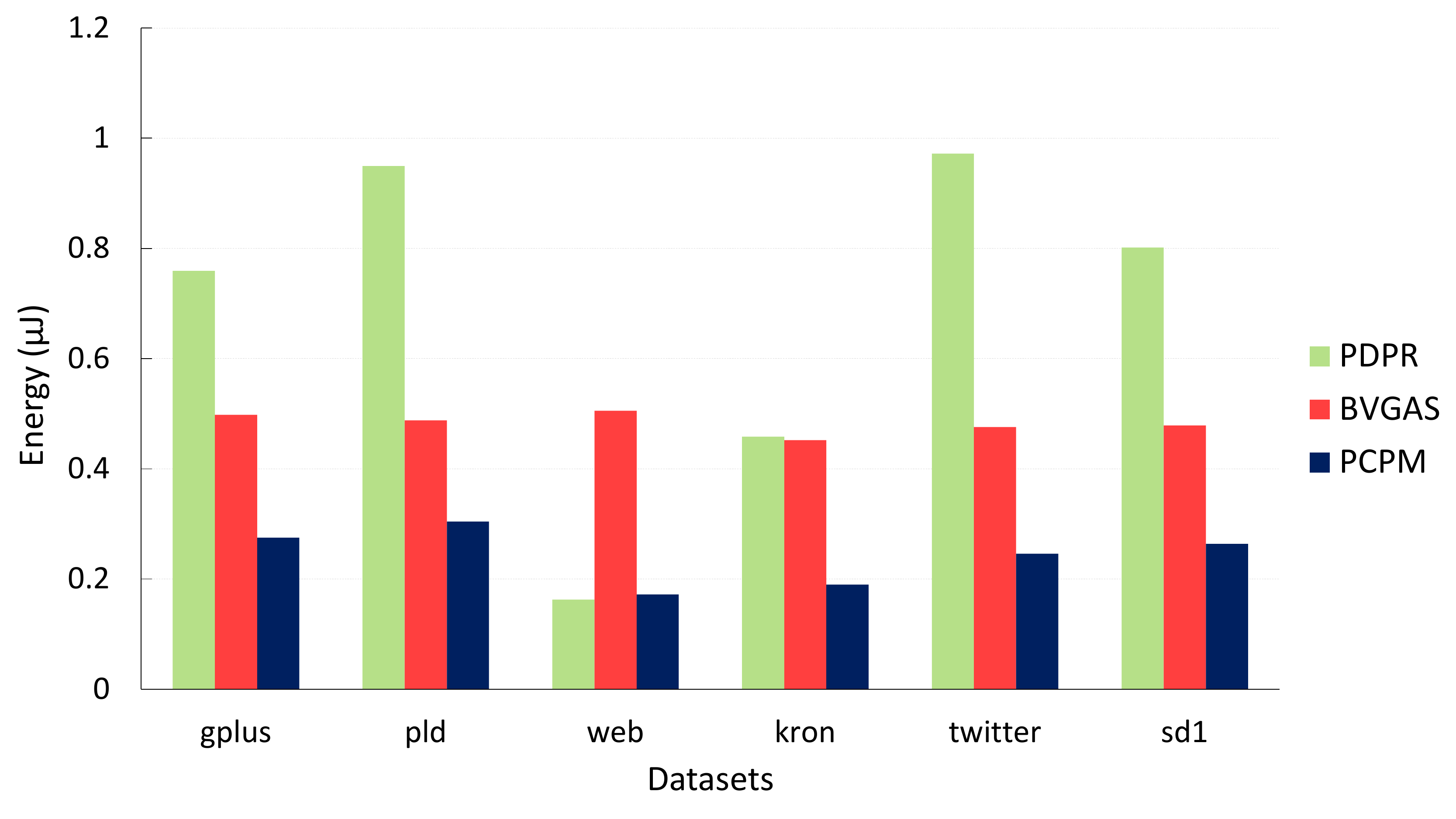}
	\caption{DRAM energy consumption per edge. PCPM benefits from reduced communication and random memory accesses.}
	\label{fig:memEnergy}
\end{figure}

The reduced communication and streaming access patterns in PCPM also enhance its energy efficiency resulting in lower $\mu J$/edge consumption as compared to BVGAS and PDPR, as shown in fig.~\ref{fig:memEnergy}. Energy efficiency is important from an eco-friendly computing perspective as highlighted by the Green Graph500 benchmark~\cite{green500}. 

\noindent\textbf{Effect of Locality:}
To assess the impact of locality on different methodologies, we relabel the nodes in our graph datasets using the GOrder~\cite{gorder} algorithm. We refer to the original node labeling in graph as \textit{Orig} and GOrder labeling as simply \textit{GOrder}. \textit{GOrder} increases spatial locality by placing nodes with common in-neighbors closer in the memory. As a result, outgoing edges of the nodes tend to be concentrated in few partitions which increases the compression ratio $r$ as shown in table~\ref{table:compRatio}. However, the \textit{web} graph exhibits near optimal compression~($r=8.4$) with \textit{Orig} and does not show improvement with \textit{GOrder}. 

Table~\ref{table:localityComm} shows the impact of \textit{GOrder} on DRAM communication. As expected, BVGAS communicates a constant amount of data for a given graph irrespective of the labeling scheme used. On the contrary, memory traffic generated by PDPR and PCPM decreases because of reduced $c_{mr}$ and increased $r$, respectively. These observations are in complete accordance with the performance models discussed in section~\ref{sec:commModel}. The effect on PCPM is not as drastic as PDPR because after $r$ becomes greater than a threshold, PCPM communication decreases slowly as shown in fig.~\ref{fig:commModel}. Nevertheless, for almost all of the datasets, the net data transferred in PCPM is remarkably lesser than both PDPR and BVGAS for either of the vertex labelings.
\begin{table}[]
	\centering
	\caption{DRAM data transfer per iteration~(in GB). PDPR and PCPM benefit from optimized node labeling}	%\vspace{-1mm}
	\label{table:localityComm}
	\resizebox{\linewidth}{!}{%	
	\begin{tabular}{ccccccc}
		& \multicolumn{2}{c}{\textbf{PDPR}}                                         & \multicolumn{2}{c}{\textbf{BVGAS}}                                        & \multicolumn{2}{c}{\textbf{PCPM}}                                         \\ \hline
		\multicolumn{1}{|c|}{\textbf{Dataset}} & \multicolumn{1}{c}{\textit{Orig}} & \multicolumn{1}{c|}{\textit{GOrder}} & \multicolumn{1}{c}{\textit{Orig}} & \multicolumn{1}{c|}{\textit{GOrder}} & \multicolumn{1}{c}{\textit{Orig}} & \multicolumn{1}{c|}{\textit{GOrder}} \\ \hline
		\multicolumn{1}{|c|}{gplus}            & \multicolumn{1}{c}{13.1}          & \multicolumn{1}{c|}{7.4}             & \multicolumn{1}{c}{9.3}           & \multicolumn{1}{c|}{9.3}             & \multicolumn{1}{c}{6.6}           & \multicolumn{1}{c|}{5.1}             \\ \hline
		\multicolumn{1}{|c|}{pld}              & \multicolumn{1}{c}{24.5}          & \multicolumn{1}{c|}{10.7}            & \multicolumn{1}{c}{12.6}          & \multicolumn{1}{c|}{12.5}            & \multicolumn{1}{c}{9.4}           & \multicolumn{1}{c|}{6.1}             \\ \hline
		\multicolumn{1}{|c|}{web}              & \multicolumn{1}{c}{7.5}           & \multicolumn{1}{c|}{7.6}             & \multicolumn{1}{c}{21.6}          & \multicolumn{1}{c|}{21.3}            & \multicolumn{1}{c}{8.5}           & \multicolumn{1}{c|}{8.4}             \\ \hline
		\multicolumn{1}{|c|}{kron}             & \multicolumn{1}{c}{18.1}          & \multicolumn{1}{c|}{10.8}            & \multicolumn{1}{c}{19.9}          & \multicolumn{1}{c|}{19.5}            & \multicolumn{1}{c}{10.4}          & \multicolumn{1}{c|}{7.5}             \\ \hline
		\multicolumn{1}{|c|}{twitter}          & \multicolumn{1}{c}{68.2}          & \multicolumn{1}{c|}{31.6}            & \multicolumn{1}{c}{28.8}          & \multicolumn{1}{c|}{28.2}            & \multicolumn{1}{c}{19.4}          & \multicolumn{1}{c|}{13.4}            \\ \hline
		\multicolumn{1}{|c|}{sd1}              & \multicolumn{1}{c}{65.1}          & \multicolumn{1}{c|}{23.8}            & \multicolumn{1}{c}{37.8}          & \multicolumn{1}{c|}{37.8}            & \multicolumn{1}{c}{26.9}          & \multicolumn{1}{c|}{15.6}            \\ \hline
	\end{tabular}
}
%\vspace{-1mm}
\end{table}

\begin{figure}[htbp]
	\includegraphics[width=\linewidth]{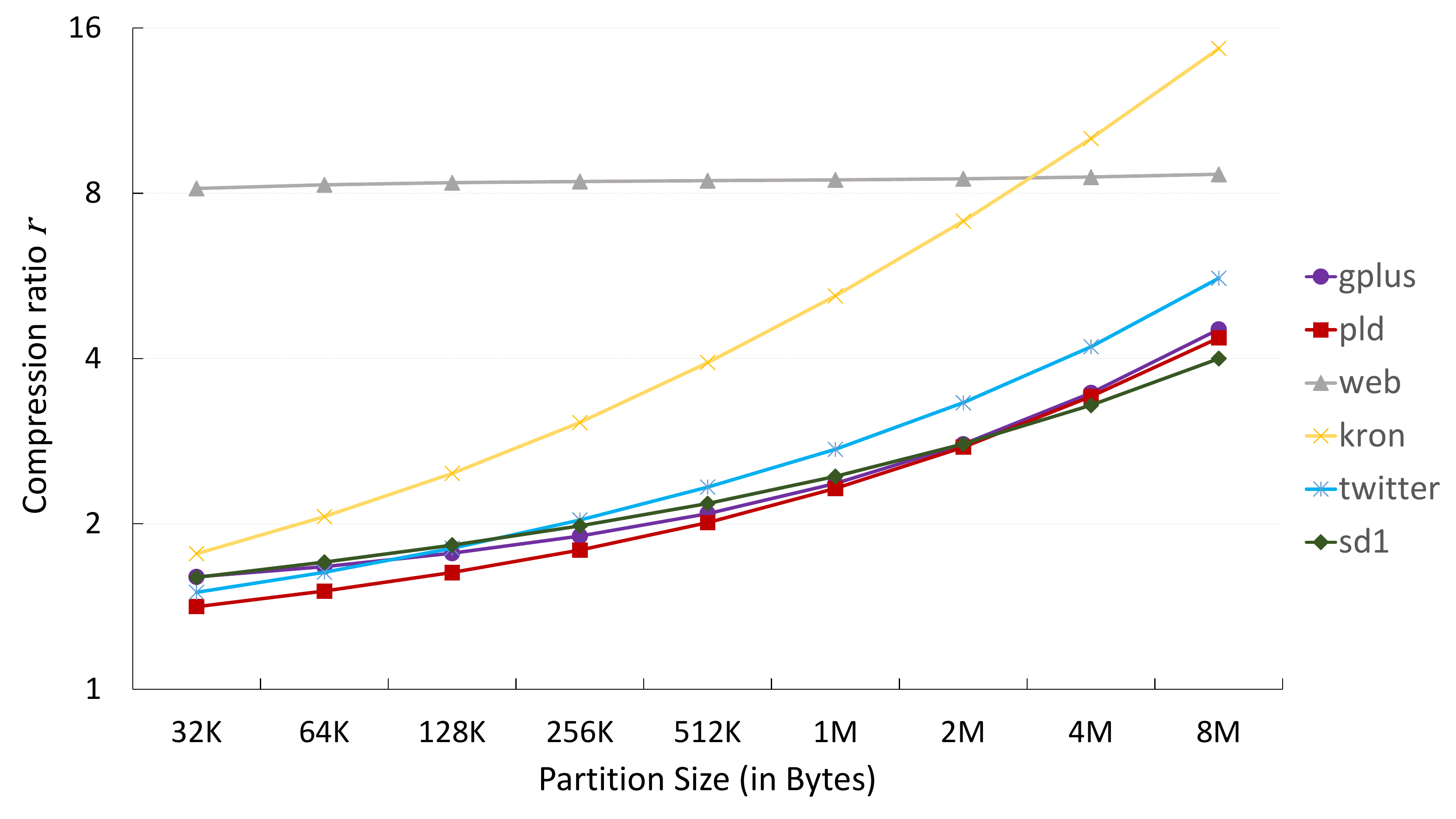}
	\caption{Compression ratio increases with partition size}
	\label{fig:compRatioVsSize}
	\vspace{-4mm}
\end{figure}

\subsubsection{PCPM Design Space Exploration}\label{sec:paramDeterm}
As we increase the size of partition, the neighbors of each node are forced to fit in fewer partitions resulting in better compression as shown in fig.~\ref{fig:compRatioVsSize}. The \textit{web} graph is an exception for which the $r$ value remains almost constant because its node labeling provides high spatial locality and close to optimal compression even for small partition sizes. The \textit{kron} dataset exhibits larger compression than other graphs because of high edge density.

\begin{figure}[htbp]
	%\vspace{-1mm}
	\includegraphics[width=\linewidth]{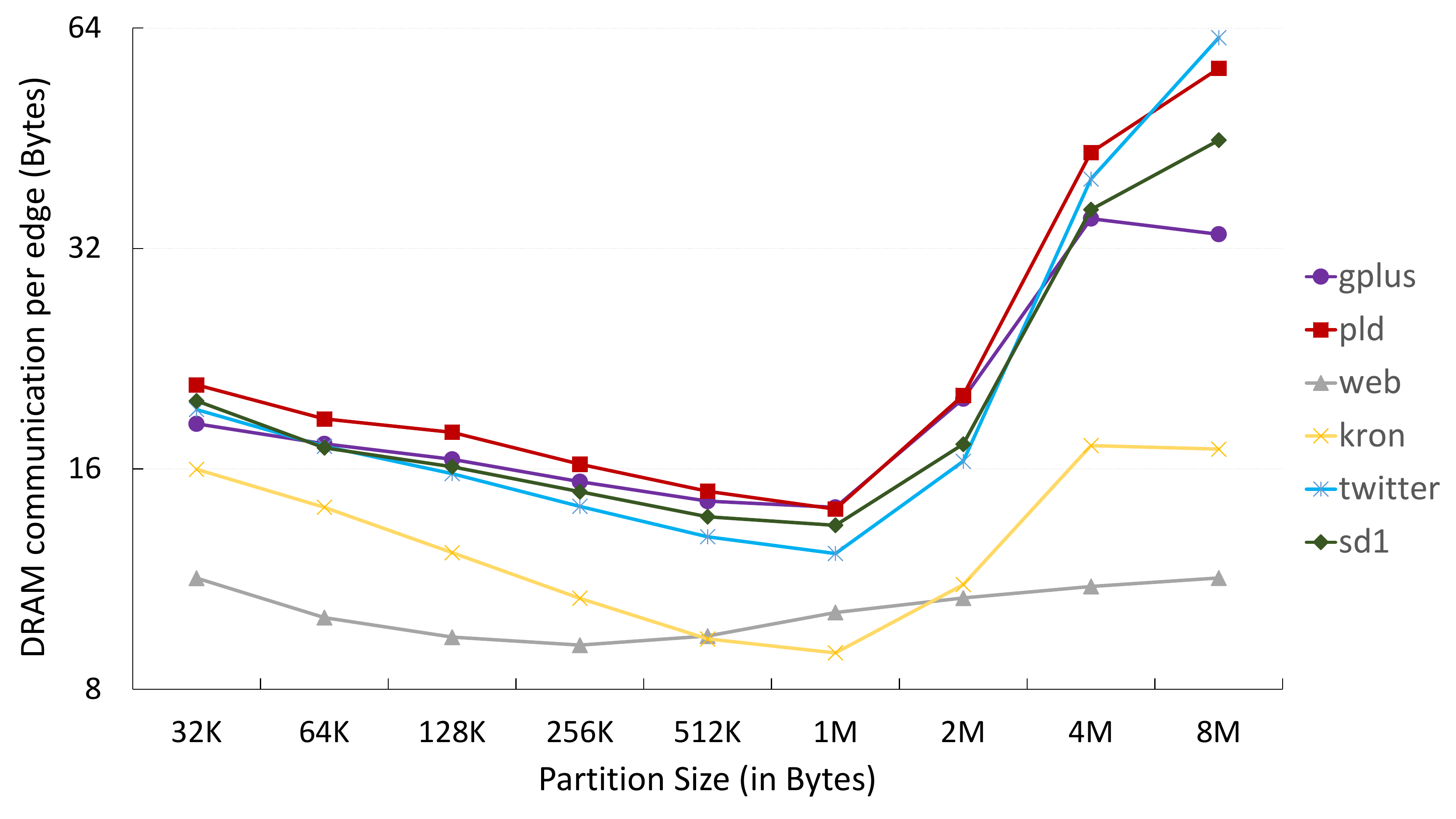}
	\caption{Impact of partition size on communication volume. Very large partitions result in cache misses and increased DRAM traffic.}
	\label{fig:trafficVsSize}%\vspace{-2mm}
\end{figure}

A direct consequence of increase in $r$ is observed in the amount of DRAM communication which reduces as we increase the partition size~(fig.~\ref{fig:trafficVsSize}). However, if the partition becomes more than the cache capacity, cache is unable to accommodate all the nodes of a partition resulting in cache misses. This drastically increases the main memory traffic as for each cache miss, one complete cache line is transferred from DRAM.

\begin{figure}[htbp]
%	%\vspace{-1mm}
	\includegraphics[width=\linewidth]{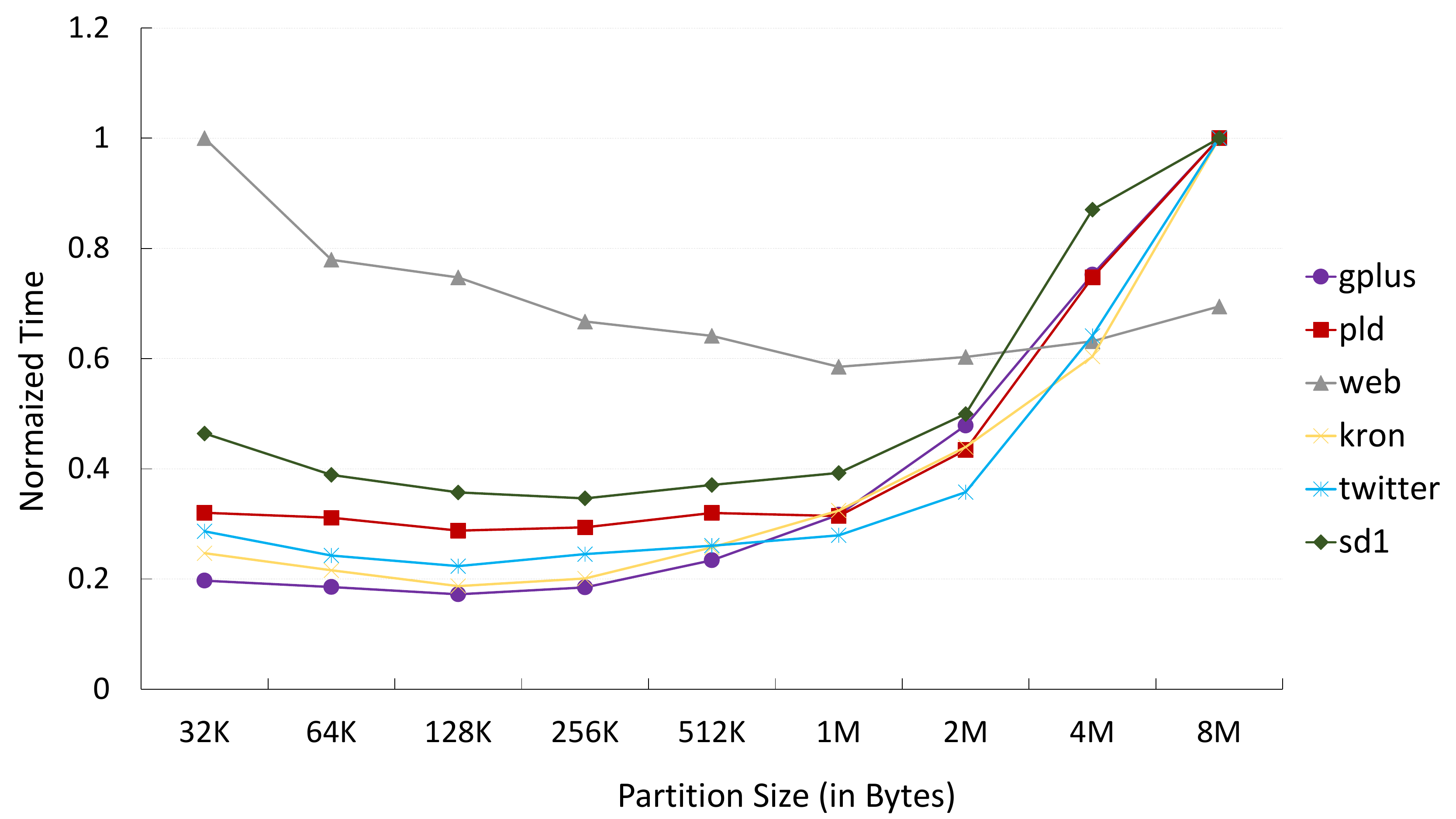}
	\caption{Impact of partition size on execution time.}
	\label{fig:timeVsSize}%\vspace{-2mm}
\end{figure}

The execution time~(fig.~\ref{fig:timeVsSize}) also benefits from communication reduction and is penalized by cache misses for large partitions. Note that for partition sizes $>256$~KB and $<=1$~MB, communication volume decreases but execution time increases. This is because in this range, many requests are served from the larger shared L3 cache which is slower than the private L1 and L2 caches. This phenomenon decelerates the computation but does not add to DRAM traffic. 
%\vspace{-1mm}

\begin{figure}
	\includegraphics[width=\linewidth]{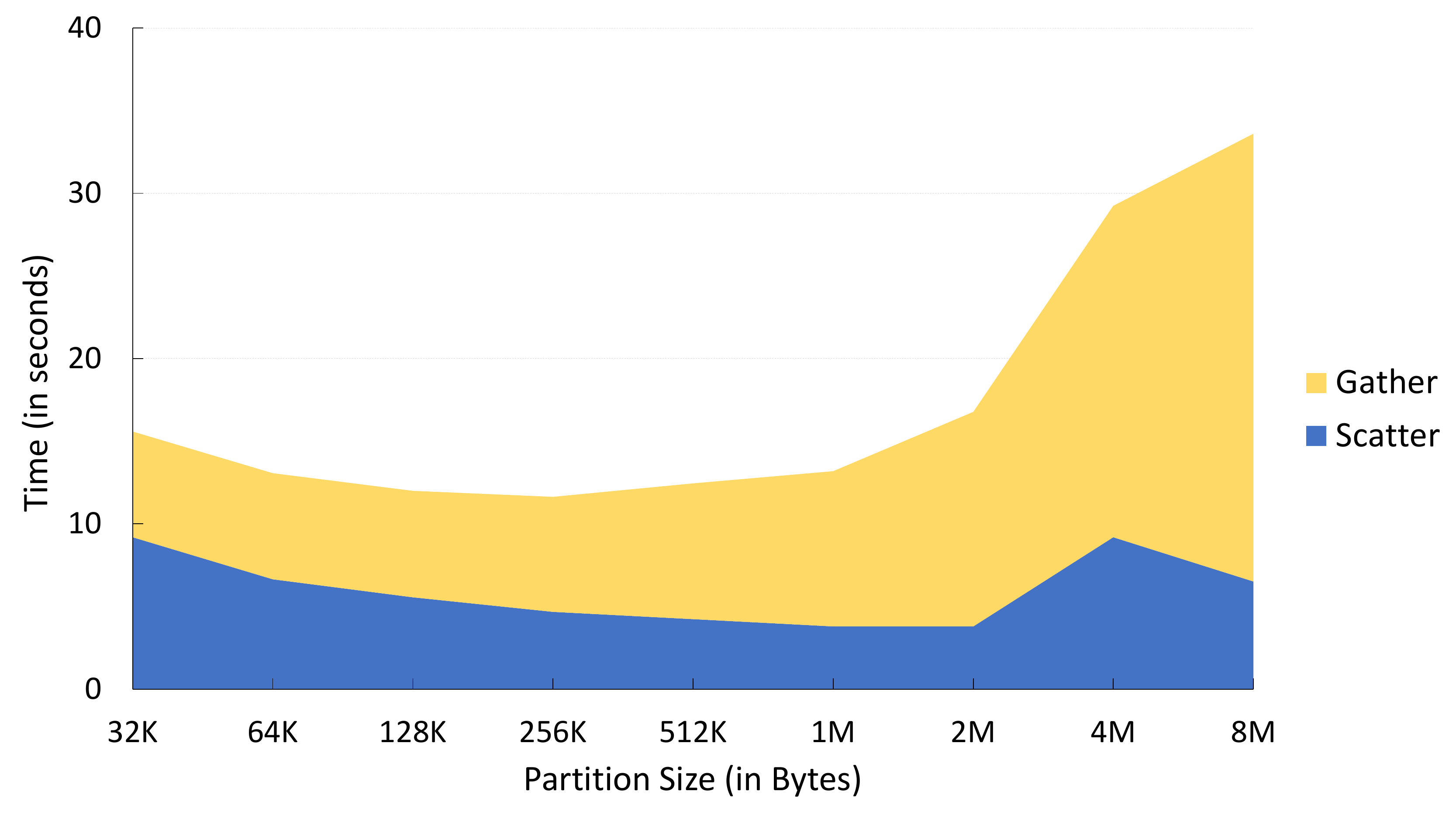}
	\caption{Time consumption of scatter and gather phases for \textit{sd1} graph. Both the phases execute fastest for partition size$=256$~KB}
	\label{fig:scatterGather}
\end{figure}

Partition size represents an important trade off in PCPM. Large partitions result in better compression but poor locality for random accesses to nodes within the partition. Fig.~\ref{fig:scatterGather} shows the effect of altering partition size separately on scatter and gather phases for PageRank computation on \textit{sd1} dataset. Both scatter and gather phases benefit from higher compression as partition size increases. However, gather phase performance saturates early because its memory accesses are proportional to $1 + \nicefrac{1}{r}$, as compared to scatter phase where accesses are proportional to \nicefrac{1}{r}. However, in both the phases, nodes within a partition are randomly accessed and hence, the performance declines if partition size grows beyond what fits in cache. Based on our observations, we chose the $256$~KB as the optimal partition size for our platform.

%\vspace{-3mm}
\subsubsection{Pre-processing Time}
We assume that adjacency matrix in CSR and CSC format is available and hence, PDPR does not need any pre-processing. Both BVGAS and PCPM however, require a beforehand computation of bin size and write offsets incurring non-zero pre-processing time as shown in table~\ref{table:preProcessing}. In addition, PCPM also constructs the PNG layout. Fortunately, the computation of write offsets is merged with PNG construction~(section~\ref{sec:PNG}) to reduce the overhead. The pre-processing time is lesser than execution time of even a single iteration~(table~\ref{table:execTime}) and gets easily amortized over multiple iterations of PageRank. 
%\vspace{-1mm}
\begin{table}[htbp]
	%\vspace{-1mm}
	\centering
	\caption{Pre-processing time of different methodologies. PNG construction increases the overhead of PCPM}
	\label{table:preProcessing}	
	\resizebox{0.7\linewidth}{!}{%	
	\begin{tabular}{|c|c|c|c|}
		\hline
		\textbf{Dataset} & \textbf{PCPM} & \textbf{BVGAS} & \textbf{PDPR} \\ \hline
		gplus            & 0.25s         & 0.1s           & 0s \\ \hline
		pld              & 0.32s         & 0.15s          & 0s \\ \hline
		web              & 0.26s         & 0.18s          & 0s \\ \hline
		kron             & 0.43s         & 0.22s          & 0s \\ \hline
		twitter          & 0.7s          & 0.27s          & 0s \\ \hline
		sd1              & 0.95s         & 0.32s          & 0s \\ \hline
	\end{tabular}
	}
\end{table}
%%\vspace{-1mm}
\section{Conclusion and Future Work}\label{sec:conclusion}%\vspace{-1mm}
In this paper, we formulated a Partition-Centric Processing Methodology~(PCPM) that perceives a graph as a set of links between nodes and partitions instead of nodes and their individual neighbors. We presented several features of this abstraction and developed system level optimizations to exploit them. 
%The idea for a new layout originated when we were trying to relax the Vertex-centric programming constraint that all outgoing edges of a node should be traversed consecutively. The additional freedom arising from treating edges individually allowed us to group edges in a way that eliminates random memory accesses.
% to sustain $>75\%$ of peak memory bandwidth. 
%We merge several pre-processing steps to reduce cost of computing PNG and show that its benefits far exceed the pre-processing overhead. 

We developed a novel PNG data layout for efficient processing using PCPM. The idea for this layout originated when we were trying to relax the Vertex-centric programming constraint that all outgoing edges of a node should be traversed consecutively. The additional freedom arising from treating edges individually allowed us to group edges in a way that eliminates random memory accesses to sustain $>75\%$ of peak memory bandwidth. 
%We merge several pre-processing steps to reduce cost of computing PNG and show that its benefits far exceed the pre-processing overhead. 

We conducted extensive analytical and experimental evaluation of our approach. Using a simple index based partitioning, we observed an average $2.7\times$ speedup in execution time and $1.7\times$ reduction in DRAM communication volume over state-of-the-art.
% We show that irrespective of the graph locality and density, PCPM remains the preferred choice of PageRank implementation as it communicates the least amount of data at higher bandwidth compared to other methodologies. 
In the future, we will explore edge partitioning models~\cite{simpleEdgePartition, balancedEdgePartition} to further reduce communication and improve load balancing for PCPM. 

Although we demonstrate the advantages of PCPM on PageRank, we show that it can be easily extended to generic SpMV computation. We believe that PCPM can be an efficient programming model for other graph algorithms or graph analytics frameworks. In this context, there are many promising directions for further exploration. For instance, the streaming memory access patterns of PNG enabled PCPM are highly suitable for High Bandwidth Memory~(HBM) and disk-based systems. Exploring PCPM as a programming model for heterogenous memory or processor architectures is an interesting avenue for future work.

PCPM accesses nodes from only one graph partition at a time. Hence, G-Store's smallest number of bits representation~\cite{gstore} can be used to reduce the memory footprint and DRAM communication even further. Devising novel methods for enhanced compression can also make PCPM amenable to be used for large-scale graph processing on commodity PCs.

\begin{footnotesize}
	\paragraph{Acknowledgements:} This material is based on work supported by the Defense Advanced	Research Projects Agency (DARPA) under Contract Number FA8750-17-C-0086, National Science Foundation (NSF) under Contract Numbers CNS-1643351 and ACI-1339756 and Air Force Research Laboratory under Grant Number FA8750-15-1-0185. Any opinions, findings and conclusions or recommendations	expressed in this material are those of the authors and do not necessarily reflect the views of DARPA, NSF or AFRL. The U.S. Government is authorized to reproduce and distribute reprints for Government purposes notwithstanding any copyright notation here on.
\end{footnotesize}
%\newpage
%\begin{spacing}{0.9}
\bibliographystyle{acm}

\bibliography{bibliography}
%\bibliography{bibliography}
%\end{spacing}

%\bibliography{bibliography} 

\end{document}